\documentclass[%
 reprint,
superscriptaddress,
 amsmath,amssymb,
 aps,
 prl,
floatfix,
]{revtex4-2}
\usepackage{graphicx}% Include figure files
\usepackage{dcolumn}% Align table columns on decimal point
\usepackage{bm}% bold math
\begin{document}

%\preprint{APS/123-QED}

\title{Colossal Spontaneous Hall Effect and Emergent Magnetism in \\ KTaO$_3$ Two-Dimensional Electron Gases}

\author{Patrick W. Krantz}
\affiliation{Department of Physics, Northwestern University, Evanston, Illinois. 60208, USA}%
\author{Alexander Tyner}%
\affiliation{Graduate Program for Applied Physics, Northwestern University, Evanston, Illinois. 60208, USA}%
\author{Pallab Goswami}
\affiliation{Department of Physics, Northwestern University, Evanston, Illinois. 60208, USA}%
\affiliation{Graduate Program for Applied Physics, Northwestern University, Evanston, Illinois. 60208, USA}%
\author{Venkat Chandrasekhar}
\affiliation{Department of Physics, Northwestern University, Evanston, Illinois. 60208, USA}%
\affiliation{Graduate Program for Applied Physics, Northwestern University, Evanston, Illinois. 60208, USA}%

\date{\today}% It is always \today, today,

\begin{abstract}
There has been intense recent interest in the two-dimensional electron gases (2DEGs) that form at the surfaces and interfaces of KTaO$_3$ (KTO), with the discovery of superconductivity at temperatures significantly higher than those of similar 2DEGs based on SrTiO$_3$ (STO). Here we demonstrate that KTO 2DEGs fabricated under conditions that suppress the superconductivity show a large spontaneous Hall effect at low temperatures. The transverse response is asymmetric in an applied perpendicular magnetic field and becomes hysteretic at millikelvin temperatures.  The hysteresis is due to long range magnetic order arising from local Ta$^{4+}$ moments. However, the most striking features of the data are the asymmetry of the transverse response and the large spontaneous transverse resistance at zero field, which can be a significant fraction of the longitudinal resistance and depends on crystal orientation. Both effects are due to the presence of a dominant contribution to the transverse response that is symmetric in perpendicular field, suggesting that its origin is topological in nature.  We argue that this contribution arises from Berry curvature dipoles coupled with nonequilibrium conditions induced by the measuring current.
\end{abstract}

\maketitle

Two-dimensional electron gases (2DEGs) in complex oxide heterostructures have seen a great deal of interest in the past two decades due to the variety of phenomena they exhibit. The most studied of these 2DEGs are those formed at the interface and surface of SrTiO$_3$ (STO) \cite{gariglio_research_2016, huang_interface_2018, pai_physics_2018}.  The phenomena seen in STO-based heterostructures include gate-tunable conductivity \cite{thiel_tunable_2006, cen_nanoscale_2008}, superconductivity \cite{reyren_superconducting_2007}, spin-orbit interactions \cite{caviglia_tunable_2010,ben_shalom_tuning_2010}, spin-polarized transport \cite{lesne_highly_2016, song_observation_2017}, quantum interference effects \cite{goswami_quantum_2016} and magnetism \cite{brinkman_magnetic_2007,dikin_coexistence_2011,bert_direct_2011,li_coexistence_2011}, among others. In the past few years, interest has turned to 2DEGs based on KTaO$_3$ (KTO)\cite{zou_latio_2015,bareille_two-dimensional_2015,wadehra_emergent_2021,gupta_ktao_2022}.  Bulk KTO is similar to STO in that is a band insulator with a large gap ($\sim$3.5 eV) \cite{wemple_transport_1965}, and it has a large dielectric constant that grows with decreasing temperature but saturates at low temperatures \cite{fujii_KTO_1976}, signaling a transition to a quantum paraelectric phase \cite{fujishita_2016}.  

Many of the studies done on STO 2DEGs mentioned above have also been performed on KTO 2DEGs.  However, there are two key differences between STO and KTO 2DEGs.  First, although superconductivity is observed in both STO and KTO 2DEGs, the superconducting transition temperatures in KTO 2DEGs are significantly higher than in STO 2DEGs \cite{liu_two_dimensional_2021}.  Second, the strength of the spin-orbit interactions in KTO is more than an order of magnitude larger than in STO \cite{zhang_high_mobility_2018,bruno_band_2019}.  While superconductivity in KTO 2DEGs has received most of the attention, the enhanced spin-orbit interactions in KTO have the potential to give rise to new effects.  Here we show that KTO 2DEGs host a large spontaneous Hall effect at low temperatures, i.e., a finite transverse resistance in zero external magnetic field.  The dependence of the transverse resistance on external magnetic fields, back gate voltage $V_g$ and the surface crystal orientation of the KTO 2DEGs coupled with density functional theory (DFT) calculations indicate that the spontaneous Hall effect arises from the nontrivial topology of the KTO surface bands. 

The devices in this study were Hall bars of length 600 $\mu$m and width 50 $\mu$m patterned by photolithography followed by successive deposition and oxidation of thin layers of Al on KTO substrates with three different surface crystal orientations: (001), (110) and (111). Prior to spinning photoresist, the substrates were cleaned using standard cleaning procedures, but were not annealed at elevated temperatures.  As a consequence, these devices do not show a superconducting transition down to our lowest measurement temperatures.  Substrates that were annealed but otherwise processed identically do go superconducting, as discussed in the Supplementary Material \cite{supplementary_reference}.  While the reason for this is not clear at present, it does enable studies of the normal state properties of KTO 2DEGs down to millikelvin temperatures, revealing surprising results.

We focus here on measurements of the longitudinal and transverse low-frequency differential resistance as a function of $V_g$, temperature $T$ and applied external field $H$ using standard low frequency ac lock-in techniques.  Importantly, the ac longitudinal and transverse voltages are measured at the same frequency as the ac current drive, with small drive amplitudes ($\sim$100 nA). A schematic of the sample can be found in Fig. \ref{fig:GateVoltageAnneal}(a), and details of the measurement can be found in the Supplementary.      

\begin{figure}
	\centering
	\includegraphics[width=\columnwidth]{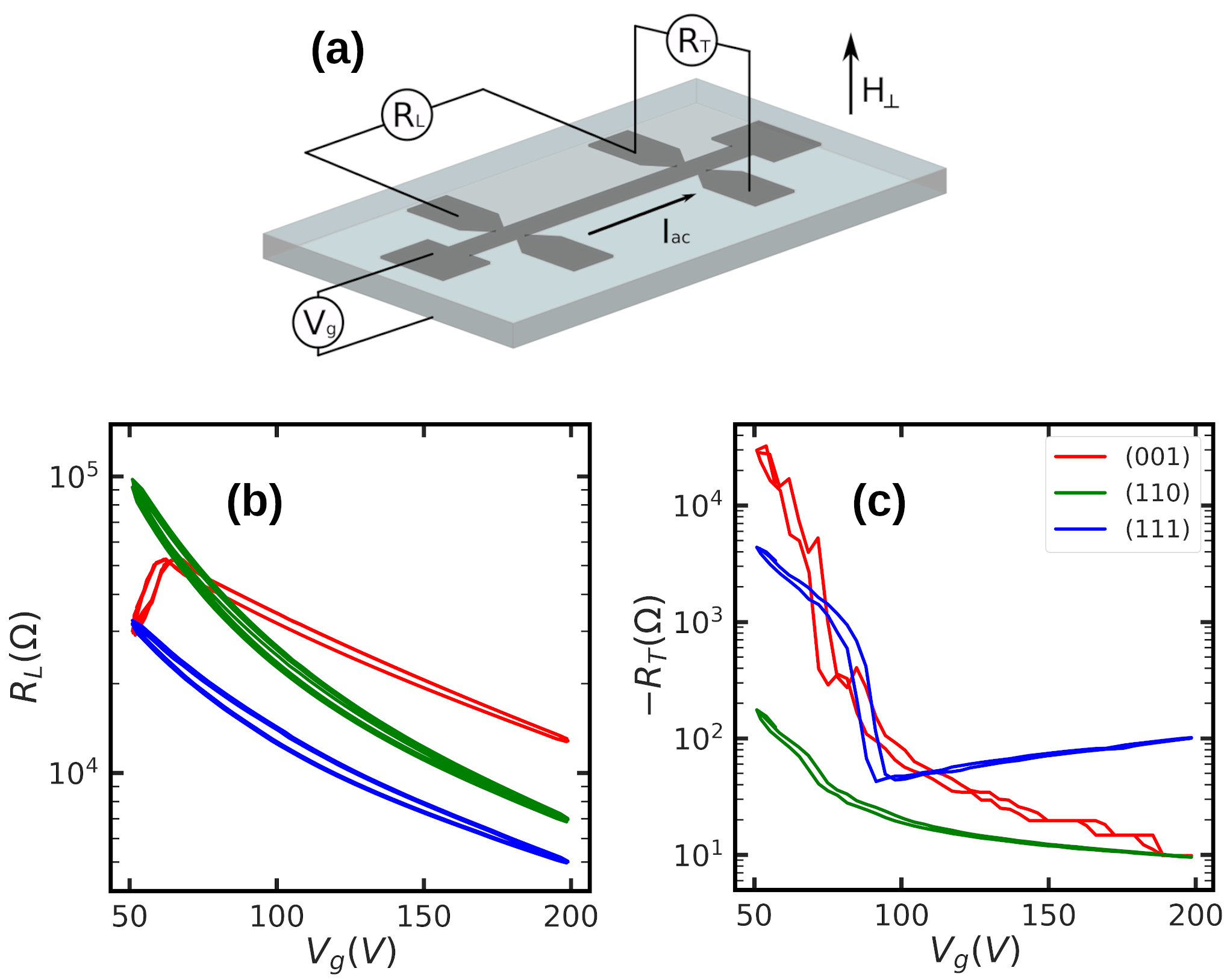}
	\caption{\label{fig:GateVoltageAnneal}  (a) Representative diagram of the sample in a Hall bar geometry, where the back gate voltage $V_g$ is applied between the conducting gas and the back of the KTaO$_3$ substrate. The current direction, the perpendicular applied magnetic field, and the contacts used for the longitudinal resistance $R_{L}$ and the transverse resistance $R_{T}$ are shown. (b) Semi-log plot of differential longitudinal resistance $R_L$ vs $V_{g}$ for each of the three crystal orientations. (c) Simultaneously measured differential transverse resistance $R_T$ vs $V_{g}$. Data were taken at 2.25K, in zero magnetic field.}
\end{figure}
      
Figure \ref{fig:GateVoltageAnneal}(b) shows the longitudinal resistance $R_L$ for $H=0$ at $T=2.25$ K as a function of $V_g$ for Hall bars on (001), (110), and (111) oriented surfaces. $R_L$ shows the same general trend for all three orientations: it increases with decreasing $V_g$ and shows hysteresis with respect to $V_g$.  Similar gate voltage dependent behavior has been reported in KTO-based heterostructures \cite{nakamura_electric_2009}, and in STO-based heterostructures \cite{davis_anisotropic_2017}.  In contrast to $R_L$, the transverse resistance $R_T$ of the KTO heterostructures shows unusual behavior.  Figure \ref{fig:GateVoltageAnneal}(c) shows the simultaneously measured $R_T$ for the three crystal orientations as a function of $V_g$ in zero external field.  At $V_g$=200 V, $R_T\sim10$ $\Omega$ for the (001) and (110) oriented Hall bars.  However, the (111) oriented Hall bar shows a much larger $R_T$ at $V_g \sim$ 200 V.  More striking is the behavior of $R_T$ as $V_g$ is decreased.  For all 3 orientations, $\lvert R_T \lvert$ increases in magnitude as $V_g$ is decreased, growing to a significant fraction of $R_L$.  In particular, there is a sharp increase in $\lvert R_T \lvert$ below $V_g \sim90$ V.  This increase is most prominent in the (111) sample, but also present in the (001) and (110) samples.  There is no corresponding feature in $R_L$, suggesting the effect does not arise from a longitudinal contribution due to device imperfections or voltage probe misalignment, but instead represents the onset of a spontaneous Hall effect. The spontaneous Hall angle $\Theta_H$, defined by $\tan(\Theta_H) = \sigma_{xy}/\sigma_{xx} = R_T/R_\square$ works out to be $\Theta_H \sim 56^\circ$ for the (111) orientation at the lowest values of $V_g$, showing that the magnitude of the effect is indeed colossal ($R_\square$ is the resistance per square). A full range of Hall angles is presented in the Supplementary Material. A similar finite $R_{T}$ is also observed in STO devices fabricated under similar conditions \cite{krantz_observation_2021}, but is smaller by an order of magnitude. 

The sweep hysteresis seen in Fig. \ref{fig:GateVoltageAnneal} is associated with glassiness in the system, and leads to a logarithmic relaxation of the measured resistance over a span of many hours, or longer, when any external parameter is changed.  To minimize the hysteresis, all the measurements reported here were taken after electrically `annealing' the sample by cycling $V_g$ repeatedly between 50 and 200 V. Even after such annealing, the resistances of the devices drift as a function of time.  Consequently, the MR data have been corrected for this drift as first described by Biscaris \textit{et al.} for STO based devices \cite{biscaras_limit_2015}. This correction is described in detail in the Supplementary Material. 

\begin{figure}
	\begin{center}
	  \includegraphics[width=0.9\columnwidth]{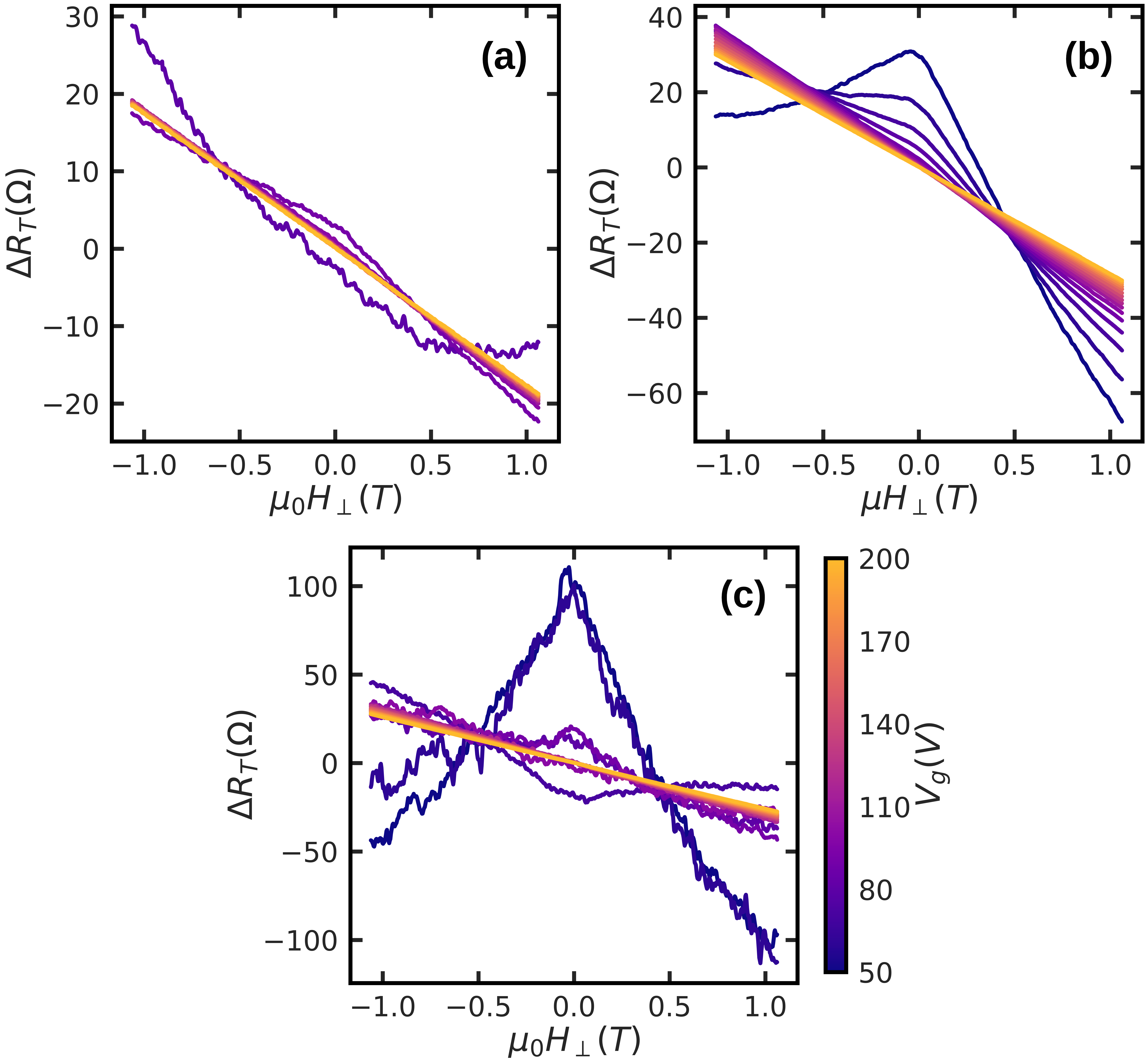}
    \end{center}
	\caption{\label{fig:HallEffect2K} Change in transverse resistance, $\Delta R_T$, as a function of a field $\mu_{0} H_{\perp}$ applied normal to the plane of the Hall Bar for the (a) (001), (b) (110) and (c) (111) oriented samples respectively. Traces for $V_g$ ranging from 50 to 200 V are shown. }
\end{figure}

\begin{figure}
	\begin{center}
	  \includegraphics[width=\columnwidth]{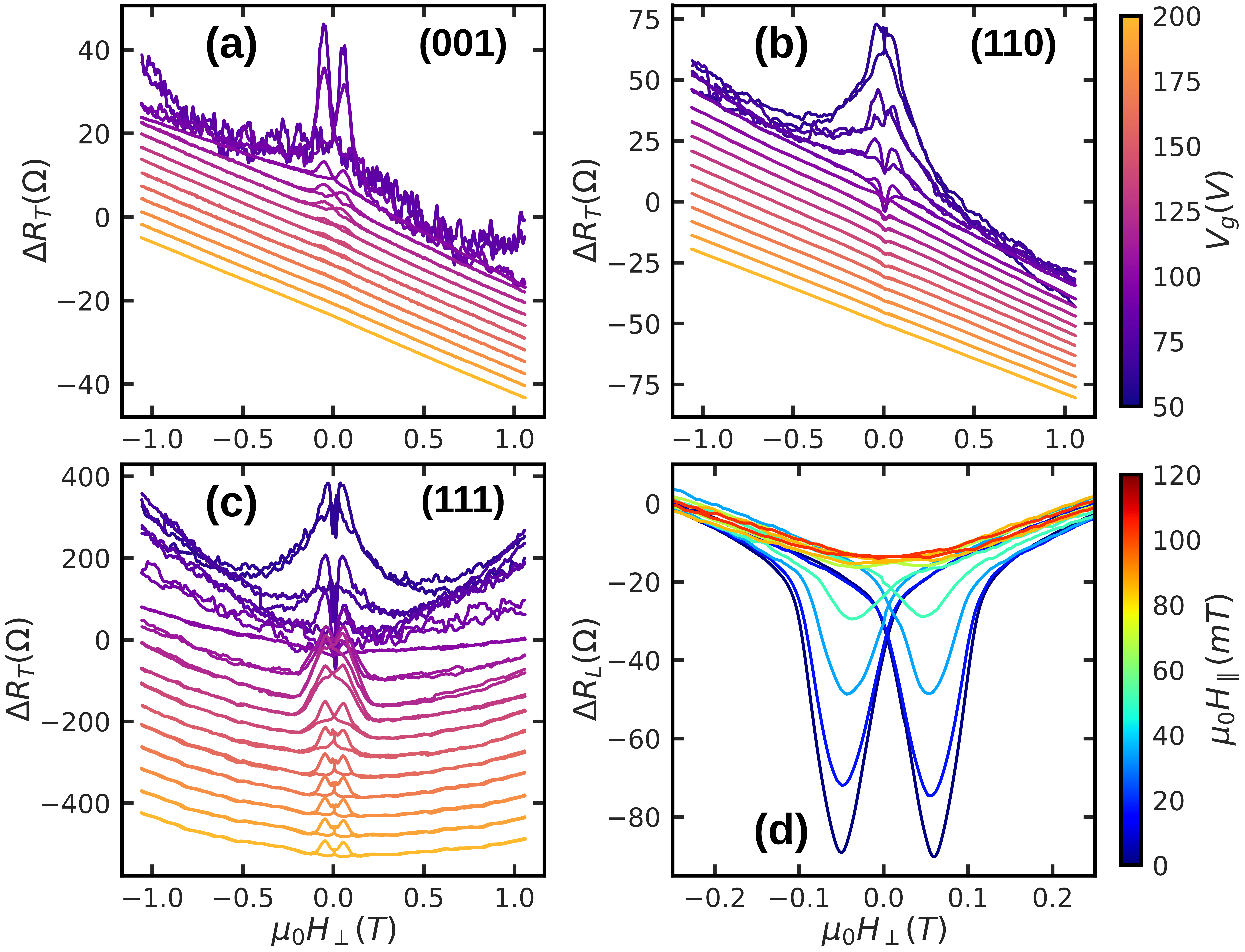}
    \end{center}
	\caption{\label{fig:HallEffectmK} (a-c) Change in transverse resistance, $\Delta R_T$, as a function of $\mu_{0} H_{\perp}$ for the (a) (001), (b) (110), and (c) (111) oriented samples, respectively, at $T\sim30$ mK. (d) Suppression of the hysteretic peaks in longitudinal resistance by application of a field $\mu_{0} H_{\parallel}$ parallel to the plane of the Hall bar, measured for the (001) oriented sample at $V_g=80$ V.}
\end{figure}

We now discuss the transverse and longitudinal magnetoresistance (MR) of the devices.  While anomalous behavior is seen in both $R_T(H)$ and $R_L(H)$, we focus on $R_T(H)$ as it is easier to separate the anomalous contributions from the contribution due to the ordinary Hall effect. Figure \ref{fig:HallEffect2K}(a-c) shows $R_T$ for the (001), (110) and (111) oriented devices as a function of field $H_\perp$ applied perpendicular to the sample at different values of $V_g$ for $T\sim$ 2.25 K (corresponding data for $\sim$5 K are shown in the Supplementary).  In order to be able to show data for all values of $V_g$ on the same graph, the respective zero-field value of $R_T$, which varies over a wide range as shown in Fig. \ref{fig:GateVoltageAnneal}, has been subtracted from each curve. At large $V_g$, $R_T (H_\perp)$ for all three  orientations is linear, with a slope corresponding to an electron density of $n\sim 4-10 \times 10^{13}/$cm$^2$ based on a fit to a single band model, comparable to densities obtained by other groups \cite{qiao_gate_2021,liu_two_dimensional_2021}.  As $V_g$ is decreased, $R_T(H_\perp)$ becomes increasingly asymmetric in $H_\perp$, with the slope of $R_T(H_\perp)$ for $H_\perp>0$ becoming different from the slope for $H_\perp<0$. In addition, the (001) and (111) devices show a change in curvature of $R_T(H_\perp)$ that occurs around $V_g\sim 90$ V (data on a finer voltage scale around this value of $V_g$ is shown in Supplementary Materials).  This change occurs at approximately the same $V_g$ below which the sharp increase in $R_T$ is seen in Fig. \ref{fig:GateVoltageAnneal}(c). At lower temperatures, the asymmetries become stronger and an additional feature is seen: the transverse response becomes hysteretic in the applied field.  This can be seen in Figs. \ref{fig:HallEffectmK}(a-c), which show the transverse response for the (001), (110), and (111) oriented devices respectively at $\sim$30 mK.

\begin{figure}
	\begin{center}
	  \includegraphics[width=\columnwidth]{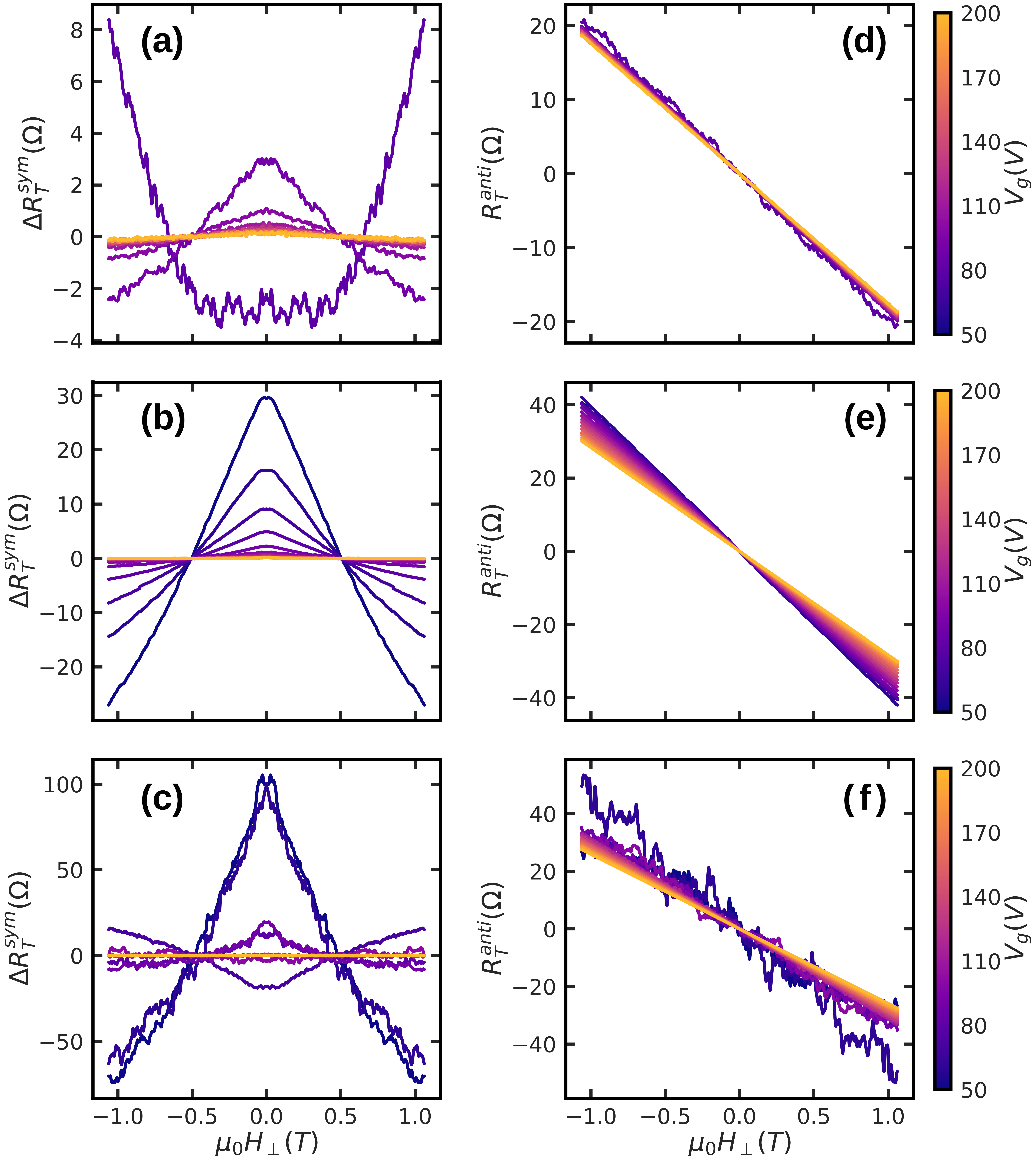}
    \end{center}
	\caption{\label{fig:AntiSymm} (a-c) Symmetric components of the $\Delta R_T$ traces for the (001), (110), and (111) oriented substrates respectively, extracted from the data shown in Fig. \ref{fig:HallEffect2K} (a-c). Corresponding antisymmetric components are shown in (d-f), from which carrier concentrations were calculated (see Supplementary Materials).}
\end{figure}

Three different contributions to the transverse response can be identified, which become clearer if we decompose the Hall response into its symmetric and antisymmetric components.  This decomposition is shown in Fig. \ref{fig:AntiSymm}(a-f) for the data shown in Fig. \ref{fig:HallEffect2K}, which was measured at 2.25 K.
The first contribution is just the ordinary Hall effect, which is the part of the transverse response that is antisymmetric in field. This is shown in Figs. \ref{fig:AntiSymm}(d), (e) and (f) for the (001), (110), and (111) devices, respectively.  The slope of this linear component for the (110) and (111) devices is roughly twice that for the (001) device, but the slopes increase in magnitude for all three orientations with decreasing $V_g$, as expected for an electron-like conducting gas. 

The second contribution is the hysteresis observed at lower temperatures.  Hysteretic behavior in the MR is usually associated with long range ferromagnetic order in the system that interacts with the conduction electrons, with the fields at which the hysteresis occurs corresponding to the coercive field of the ferromagnet.  To demonstrate that the hysteresis we observe does indeed arise from magnetic order, Fig. \ref{fig:HallEffectmK}(d) shows the longitudinal MR of the (001) sample in perpendicular field at $V_g=80$ V, with progressively larger values of an in-plane field $H_\parallel$ applied.  It can be seen that by $H_\parallel\sim$ 100 mT, the hysteresis in $H_\perp$ is completely suppressed, suggesting that once the ferromagnetic moments are fully aligned in-plane, they no longer can give rise to hysteretic MR. 

The presence of ferromagnetic order in our samples is surprising. In contrast to the EuO/KTO samples studied by other groups where hysteretic behavior has been reported \cite{zhang_high_mobility_2018}, there is no intrinsically magnetic component in our samples. Similar emergent hysteresis has been observed in STO 2DEGS \cite{brinkman_magnetic_2007,dikin_coexistence_2011}. In the case of STO, it was proposed that Ti$^{3+}$ ions could form local moments due to unpaired electrons in an otherwise unoccupied $d$-orbital \cite{pavlenko_magnetic_2012}.  Our density functional theory (DFT) calculations show that in the presence of oxygen vacancies, Ta$^{4+}$ ions can similarly form local moments that give rise to a surface magnetism (see Supplementary Materials). Thus, we propose that the hysteresis observed in our KTO devices is due to local Ta$^{4+}$ that order at lower temperatures. It is well known that such localized moments can give rise to anisotropic magnetoresistance (AMR) \cite{mcguire_anisotropic_1975} and an anomalous Hall effect (AHE) \cite{nagaosa_anomalous_2010}.  In conventional ferromagnets, for example, this leads to a contribution to the Hall effect that is proportional to the magnetization (AHE), and a contribution to the longitudinal MR that depends on the angle between the magnetization $\mathbf{M}$ and the current (AMR). 

\begin{figure}
	\centering
	\includegraphics[width=\columnwidth]{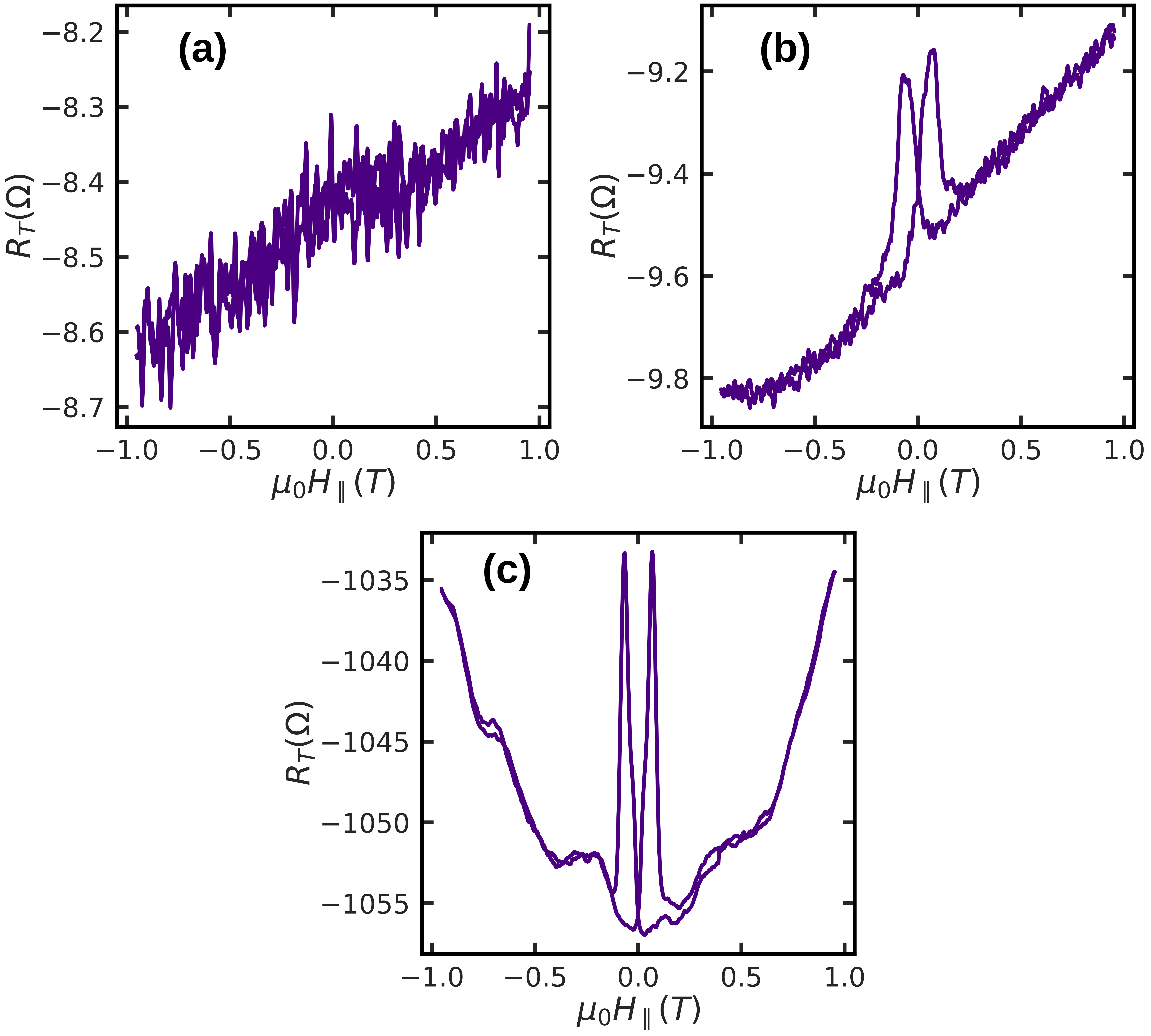}
	\caption{\label{fig:ParallelField} Parallel field Hall curves at $\boldsymbol{\sim}$30mK and $V_{g}$ = 200V. (a) shows a very small planar Hall effect of the (001) sample while (b) shows a larger background planar Hall for the (110) sample with the emergence of hysteretic peaks. (c) shows the response of the (111) sample, with both hysteretic peaks at low fields and background oscillations at larger fields.}
\end{figure}

The third contribution to the transverse response is the part that is symmetric in field, shown in Figs. \ref{fig:AntiSymm} (a), (b), and (c) for the (001), (110), and (111) orientations, respectively. The data considered here are reported at 2.25 K, above the temperature at which the hysteretic behavior emerges. This symmetric component is the dominant contribution to the transverse response at lower values of $V_g$ and $T$. We remind the reader that Fig. \ref{fig:AntiSymm} shows only the small variation in the magnitude of the symmetric component that this variation is on top of the larger zero field $R_T$ shown in Fig. \ref{fig:GateVoltageAnneal}(c). 

Observation of a transverse resistance $R_T$ is thought to require time reversal symmetry (TRS) breaking, typically introduced by applying an external magnetic field.  In this case, fundamental symmetries dictate that $R_T$ be antisymmetric in the external field, as is well known from the ordinary Hall effect, implying that $R_T$ vanishes in zero field.  One might observe a finite $R_T$ at zero field if there are intrinsic sources of TRS breaking like a finite magnetic moment, but this would result in hysteresis associated with magnetization switching superposed on an antisymmetric background.  Importantly, the zero field value in this case would still swtich sign as the magnetization reversed.  The most direct evidence that the symmetric component does not arise from an intrinsic magnetization is shown in Fig. \ref{fig:ParallelField}, which shows $R_T$ as a function of an applied in-plane magnetic field $H_\parallel$.  At parallel fields much larger than required to suppress the hysteresis in perpendicular field (see Fig. \ref{fig:HallEffectmK}(d)), the overall value of $R_T$ changes by no more than a few percent.

There have been reports of spontaneous Hall effects in three-dimensional crystals with no net magnetization \cite{nakatsuji_spontaneous_2011,ueda_spontaneous_2018,dzsaber_giant_2021} ascribed to the Berry curvature $\boldsymbol{\Omega}(\boldsymbol{k})$ of the bands and its associated anomalous velocity.  In general, in a system with TRS but broken inversion symmetry, $\boldsymbol{\Omega}(\boldsymbol{k})$ is an odd function of the crystal momentum $\boldsymbol{k}$, so that integrating over the Brillouin zone, with equal occupation of $+\boldsymbol{k}, -\boldsymbol{k}$ states, the Berry flux vanishes.  However, in a nonequilibrium situation such as in a transport measurement, the occupation of $+\boldsymbol{k}, -\boldsymbol{k}$ states is not the same, potentially generating a net transverse current from anomalous velocity arising from the Berry curvature.  This is predicted to give rise to a nonlinear Hall effect, with an ac current at frequency $f$ generating transverse Hall voltages at 0 and $2f$ \cite{sodemann_quantum_2015}, and has been reported, for example, by Dzsaber \textit{et al.} \cite{dzsaber_giant_2021} in experiments on Ce$_3$Bi$_4$Pd$_3$. However, Dzsaber \textit{et al.} also observe a dominant signal at $f$.  While a dominant signal at $f$ cannot be explained by the perturbative theory of Ref. [\citenum{sodemann_quantum_2015}], Dzaber \textit{et al.} suggest that the origin of their dominant $f$ signal is still due to Berry curvature but in a regime where a perturbative expansion cannot be used.  We believe that the spontaneous Hall effect that we observe is also due to the Berry curvature of the Rashba spin-split bands in KTO, though this large response in $f$ does not agree with the current theory of the nonlinear Hall effect.

\begin{figure}[htbp]
	\centering
	\includegraphics[width=0.6\columnwidth]{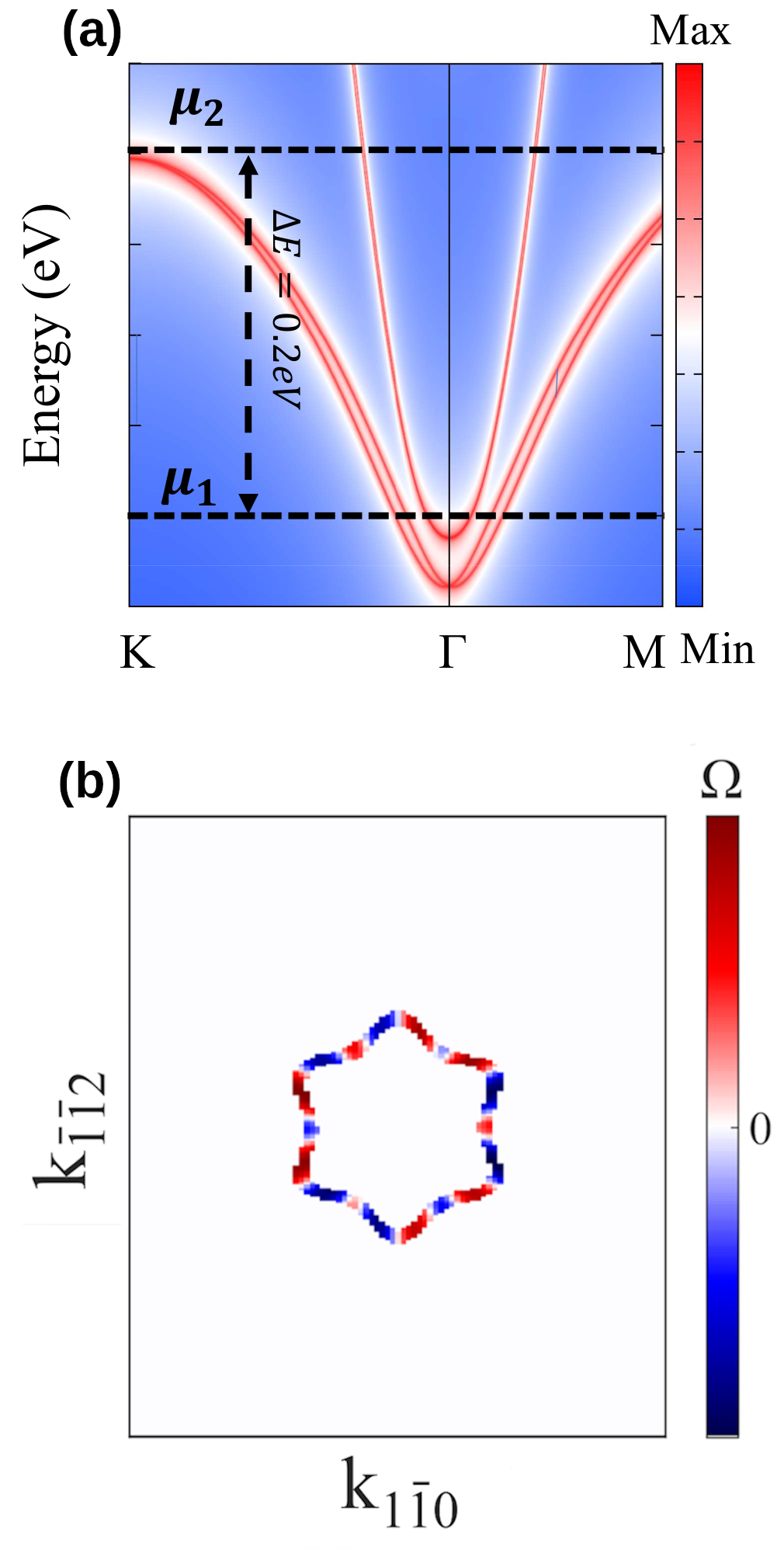}
	\caption{\label{fig:BerryDipole} (a) Spectral density on the (111) surface along high-symmetry path in the Brillouin zome. Berry curvature density is calculated at two filling fractions, $\mu_{1}$ and $\mu_{2}$. (b) Berry curvature density at low filling fraction ($\mu_{1}$). The finite Berry curvature arises when only the lowest lying Rashba split band is occupied. The cut at ($\mu_{2}$) has diminished Berry curvature density and is shown in the Supplemental Materials.}
\end{figure}

%It has been shown that in the presence of a warping term due to the $C_3$ symmetry of the (111) surface, the low-energy Rashba Hamiltonian describing the 2DEG can support a non-vanishing Berry curvature density as a function of filling \cite{lesne_designing_2022}. 

The Berry curvature density is finite only only in those regions of momentum space when a single band in the Rashba-split subspace is occupied for a specific chemical potential $\mu$: if both bands are occupied their contributions cancel,   As a result, the area in momentum space over which the Berry curvature density is finite will correlate directly with the strength of the Rashba splitting, i.e., a stronger Rashba splitting yields a large area supporting a finite Berry curvature density at a given $\mu$. KTaO$_{3}$ is known to support a significantly larger Rashba coefficient when compared to SrTiO$_{3}$ \cite{bruno_band_2019}. As a consequence, we expect the presence of a finite Berry curvature to have a more significant impact on the physical response.  The unique feature of the KTO devices is that we can change $\mu$ by changing $V_g$, thereby changing the Berry curvature of the bands and the resulting spontaneous Hall effect.

To demonstrate this, we directly calculate the Berry curvature on the (111) surface utilizing a Wannier tight binding model with a surface potential well. Details of the calculation are discussed in the Supplementary Materials. The band structure for the (111) surface is shown in Fig. \ref{fig:BerryDipole}(a).  Figure \ref{fig:BerryDipole}(b) shows the Berry curvature $\boldsymbol{\Omega}$ at a cut corresponding to the chemical potential $\mu_1$ shown in \ref{fig:BerryDipole}(a). The Berry curvature nominally has contributions from all the bands in Fig. \ref{fig:BerryDipole}(a).  However, only the contributions from the strongly Rashba-split bands is appreciable; the contribution of the lighter bands closer to the zone center is negligible and is not visible on the color scale.  If the chemical potential is raised so that the strongly Rashba-split bands are fully occupied ($\mu_2$ in Fig. \ref{fig:BerryDipole}(a)), the Berry curvature vanishes.  This agrees with our experimental observation that the spontaneous Hall effect is greatly reduced as $V_g$ is increased beyond $\sim$100 V.  Calculations for the (001) and (110) structures are discussed in the Supplementary Material. They show that the Rashba splitting is small compared to the (111), and as a consequence the Berry curvature is negligible for (001) oriented devices and small but finite for the (110) structures. These numerical results are in excellent agreement with our experimental results, and with the low-energy analysis given by Lesne \textit{et. al} \cite{lesne_designing_2022} for LAO/STO heterostructures.

In summary, we have observed a colossal spontaneous Hall effect in KTO heterostructures arising from the topological nature of its band structure.  We also observe evidence of long range magnetic order at millikelvin temperatures that arises from local Ta$^{4+}$ magnetic moments.  The dependence of these phenomenon on gate voltage makes this a promising platform for further controlled study. In particular, further experimentation would test if these effects coexists with the superconductivity observed in samples fabricated through similar processes. 

\begin{acknowledgements}
This work was supported by the U.S. Department of Energy, Basic Energy Sciences, under Award No. DE-FG02-06ER46346. This work also made use of the NUFAB facility of Northwestern University’s NUANCE Center, which has received support from the SHyNE Resource (NSF ECCS-2025633), the IIN, and Northwestern’s MRSEC program (NSF DMR-1720139). Additional equipment support was provided by DURIP grant W911NF-20-1-0066.
\end{acknowledgements}

\bibliography{KTO_bib}% Produces the bibliography via BibTeX.

\providecommand{\noopsort}[1]{}\providecommand{\singleletter}[1]{#1}%
\begin{thebibliography}{39}%
\makeatletter
\providecommand \@ifxundefined [1]{%
 \@ifx{#1\undefined}
}%
\providecommand \@ifnum [1]{%
 \ifnum #1\expandafter \@firstoftwo
 \else \expandafter \@secondoftwo
 \fi
}%
\providecommand \@ifx [1]{%
 \ifx #1\expandafter \@firstoftwo
 \else \expandafter \@secondoftwo
 \fi
}%
\providecommand \natexlab [1]{#1}%
\providecommand \enquote  [1]{``#1''}%
\providecommand \bibnamefont  [1]{#1}%
\providecommand \bibfnamefont [1]{#1}%
\providecommand \citenamefont [1]{#1}%
\providecommand \href@noop [0]{\@secondoftwo}%
\providecommand \href [0]{\begingroup \@sanitize@url \@href}%
\providecommand \@href[1]{\@@startlink{#1}\@@href}%
\providecommand \@@href[1]{\endgroup#1\@@endlink}%
\providecommand \@sanitize@url [0]{\catcode `\\12\catcode `\$12\catcode
  `\&12\catcode `\#12\catcode `\^12\catcode `\_12\catcode `\%12\relax}%
\providecommand \@@startlink[1]{}%
\providecommand \@@endlink[0]{}%
\providecommand \url  [0]{\begingroup\@sanitize@url \@url }%
\providecommand \@url [1]{\endgroup\@href {#1}{\urlprefix }}%
\providecommand \urlprefix  [0]{URL }%
\providecommand \Eprint [0]{\href }%
\providecommand \doibase [0]{https://doi.org/}%
\providecommand \selectlanguage [0]{\@gobble}%
\providecommand \bibinfo  [0]{\@secondoftwo}%
\providecommand \bibfield  [0]{\@secondoftwo}%
\providecommand \translation [1]{[#1]}%
\providecommand \BibitemOpen [0]{}%
\providecommand \bibitemStop [0]{}%
\providecommand \bibitemNoStop [0]{.\EOS\space}%
\providecommand \EOS [0]{\spacefactor3000\relax}%
\providecommand \BibitemShut  [1]{\csname bibitem#1\endcsname}%
\let\auto@bib@innerbib\@empty
%</preamble>
\bibitem [{\citenamefont {Gariglio}\ \emph {et~al.}(2016)\citenamefont
  {Gariglio}, \citenamefont {Gabay},\ and\ \citenamefont
  {Triscone}}]{gariglio_research_2016}%
  \BibitemOpen
  \bibfield  {author} {\bibinfo {author} {\bibfnamefont {S.}~\bibnamefont
  {Gariglio}}, \bibinfo {author} {\bibfnamefont {M.}~\bibnamefont {Gabay}},\
  and\ \bibinfo {author} {\bibfnamefont {J.-M.}\ \bibnamefont {Triscone}},\
  }\href {https://doi.org/10.1063/1.4953822} {\bibfield  {journal} {\bibinfo
  {journal} {APL Materials}\ }\textbf {\bibinfo {volume} {4}},\ \bibinfo
  {pages} {060701} (\bibinfo {year} {2016})}\BibitemShut {NoStop}%
\bibitem [{\citenamefont {Huang}\ \emph {et~al.}(2018)\citenamefont {Huang},
  \citenamefont {{Ariando}}, \citenamefont {Renshaw~Wang}, \citenamefont
  {Rusydi}, \citenamefont {Chen}, \citenamefont {Yang},\ and\ \citenamefont
  {Venkatesan}}]{huang_interface_2018}%
  \BibitemOpen
  \bibfield  {author} {\bibinfo {author} {\bibfnamefont {Z.}~\bibnamefont
  {Huang}}, \bibinfo {author} {\bibnamefont {{Ariando}}}, \bibinfo {author}
  {\bibfnamefont {X.}~\bibnamefont {Renshaw~Wang}}, \bibinfo {author}
  {\bibfnamefont {A.}~\bibnamefont {Rusydi}}, \bibinfo {author} {\bibfnamefont
  {J.}~\bibnamefont {Chen}}, \bibinfo {author} {\bibfnamefont {H.}~\bibnamefont
  {Yang}},\ and\ \bibinfo {author} {\bibfnamefont {T.}~\bibnamefont
  {Venkatesan}},\ }\href {https://doi.org/10.1002/adma.201802439} {\bibfield
  {journal} {\bibinfo  {journal} {Advanced Materials}\ }\textbf {\bibinfo
  {volume} {30}},\ \bibinfo {pages} {1802439} (\bibinfo {year}
  {2018})}\BibitemShut {NoStop}%
\bibitem [{\citenamefont {Pai}\ \emph {et~al.}(2018)\citenamefont {Pai},
  \citenamefont {Tylan-Tyler}, \citenamefont {Irvin},\ and\ \citenamefont
  {Levy}}]{pai_physics_2018}%
  \BibitemOpen
  \bibfield  {author} {\bibinfo {author} {\bibfnamefont {Y.-Y.}\ \bibnamefont
  {Pai}}, \bibinfo {author} {\bibfnamefont {A.}~\bibnamefont {Tylan-Tyler}},
  \bibinfo {author} {\bibfnamefont {P.}~\bibnamefont {Irvin}},\ and\ \bibinfo
  {author} {\bibfnamefont {J.}~\bibnamefont {Levy}},\ }\href
  {https://doi.org/10.1088/1361-6633/aa892d} {\bibfield  {journal} {\bibinfo
  {journal} {Reports on Progress in Physics}\ }\textbf {\bibinfo {volume}
  {81}},\ \bibinfo {pages} {036503} (\bibinfo {year} {2018})}\BibitemShut
  {NoStop}%
\bibitem [{\citenamefont {Thiel}\ \emph {et~al.}(2006)\citenamefont {Thiel},
  \citenamefont {Hammerl}, \citenamefont {Schmehl}, \citenamefont {Schneider},\
  and\ \citenamefont {Mannhart}}]{thiel_tunable_2006}%
  \BibitemOpen
  \bibfield  {author} {\bibinfo {author} {\bibfnamefont {S.}~\bibnamefont
  {Thiel}}, \bibinfo {author} {\bibfnamefont {G.}~\bibnamefont {Hammerl}},
  \bibinfo {author} {\bibfnamefont {A.}~\bibnamefont {Schmehl}}, \bibinfo
  {author} {\bibfnamefont {C.~W.}\ \bibnamefont {Schneider}},\ and\ \bibinfo
  {author} {\bibfnamefont {J.}~\bibnamefont {Mannhart}},\ }\href
  {https://doi.org/10.1126/science.1131091} {\bibfield  {journal} {\bibinfo
  {journal} {Science}\ }\textbf {\bibinfo {volume} {313}},\ \bibinfo {pages}
  {1942} (\bibinfo {year} {2006})}\BibitemShut {NoStop}%
\bibitem [{\citenamefont {Cen}\ \emph {et~al.}(2008)\citenamefont {Cen},
  \citenamefont {Thiel}, \citenamefont {Hammerl}, \citenamefont {Schneider},
  \citenamefont {Andersen}, \citenamefont {Hellberg}, \citenamefont
  {Mannhart},\ and\ \citenamefont {Levy}}]{cen_nanoscale_2008}%
  \BibitemOpen
  \bibfield  {author} {\bibinfo {author} {\bibfnamefont {C.}~\bibnamefont
  {Cen}}, \bibinfo {author} {\bibfnamefont {S.}~\bibnamefont {Thiel}}, \bibinfo
  {author} {\bibfnamefont {G.}~\bibnamefont {Hammerl}}, \bibinfo {author}
  {\bibfnamefont {C.~W.}\ \bibnamefont {Schneider}}, \bibinfo {author}
  {\bibfnamefont {K.~E.}\ \bibnamefont {Andersen}}, \bibinfo {author}
  {\bibfnamefont {C.~S.}\ \bibnamefont {Hellberg}}, \bibinfo {author}
  {\bibfnamefont {J.}~\bibnamefont {Mannhart}},\ and\ \bibinfo {author}
  {\bibfnamefont {J.}~\bibnamefont {Levy}},\ }\href
  {https://doi.org/10.1038/nmat2136} {\bibfield  {journal} {\bibinfo  {journal}
  {Nature Materials}\ }\textbf {\bibinfo {volume} {7}},\ \bibinfo {pages} {298}
  (\bibinfo {year} {2008})}\BibitemShut {NoStop}%
\bibitem [{\citenamefont {Reyren}\ \emph {et~al.}(2007)\citenamefont {Reyren},
  \citenamefont {Thiel}, \citenamefont {Caviglia}, \citenamefont {Kourkoutis},
  \citenamefont {Hammerl}, \citenamefont {Richter}, \citenamefont {Schneider},
  \citenamefont {Kopp}, \citenamefont {Rüetschi}, \citenamefont {Jaccard},
  \citenamefont {Gabay}, \citenamefont {Muller}, \citenamefont {Triscone},\
  and\ \citenamefont {Mannhart}}]{reyren_superconducting_2007}%
  \BibitemOpen
  \bibfield  {author} {\bibinfo {author} {\bibfnamefont {N.}~\bibnamefont
  {Reyren}}, \bibinfo {author} {\bibfnamefont {S.}~\bibnamefont {Thiel}},
  \bibinfo {author} {\bibfnamefont {A.~D.}\ \bibnamefont {Caviglia}}, \bibinfo
  {author} {\bibfnamefont {L.~F.}\ \bibnamefont {Kourkoutis}}, \bibinfo
  {author} {\bibfnamefont {G.}~\bibnamefont {Hammerl}}, \bibinfo {author}
  {\bibfnamefont {C.}~\bibnamefont {Richter}}, \bibinfo {author} {\bibfnamefont
  {C.~W.}\ \bibnamefont {Schneider}}, \bibinfo {author} {\bibfnamefont
  {T.}~\bibnamefont {Kopp}}, \bibinfo {author} {\bibfnamefont {A.-S.}\
  \bibnamefont {Rüetschi}}, \bibinfo {author} {\bibfnamefont {D.}~\bibnamefont
  {Jaccard}}, \bibinfo {author} {\bibfnamefont {M.}~\bibnamefont {Gabay}},
  \bibinfo {author} {\bibfnamefont {D.~A.}\ \bibnamefont {Muller}}, \bibinfo
  {author} {\bibfnamefont {J.-M.}\ \bibnamefont {Triscone}},\ and\ \bibinfo
  {author} {\bibfnamefont {J.}~\bibnamefont {Mannhart}},\ }\href
  {https://doi.org/10.1126/science.1146006} {\bibfield  {journal} {\bibinfo
  {journal} {Science}\ }\textbf {\bibinfo {volume} {317}},\ \bibinfo {pages}
  {1196} (\bibinfo {year} {2007})}\BibitemShut {NoStop}%
\bibitem [{\citenamefont {Caviglia}\ \emph {et~al.}(2010)\citenamefont
  {Caviglia}, \citenamefont {Gabay}, \citenamefont {Gariglio}, \citenamefont
  {Reyren}, \citenamefont {Cancellieri},\ and\ \citenamefont
  {Triscone}}]{caviglia_tunable_2010}%
  \BibitemOpen
  \bibfield  {author} {\bibinfo {author} {\bibfnamefont {A.~D.}\ \bibnamefont
  {Caviglia}}, \bibinfo {author} {\bibfnamefont {M.}~\bibnamefont {Gabay}},
  \bibinfo {author} {\bibfnamefont {S.}~\bibnamefont {Gariglio}}, \bibinfo
  {author} {\bibfnamefont {N.}~\bibnamefont {Reyren}}, \bibinfo {author}
  {\bibfnamefont {C.}~\bibnamefont {Cancellieri}},\ and\ \bibinfo {author}
  {\bibfnamefont {J.-M.}\ \bibnamefont {Triscone}},\ }\href
  {https://doi.org/10.1103/PhysRevLett.104.126803} {\bibfield  {journal}
  {\bibinfo  {journal} {Physical Review Letters}\ }\textbf {\bibinfo {volume}
  {104}},\ \bibinfo {pages} {126803} (\bibinfo {year} {2010})}\BibitemShut
  {NoStop}%
\bibitem [{\citenamefont {Ben~Shalom}\ \emph {et~al.}(2010)\citenamefont
  {Ben~Shalom}, \citenamefont {Sachs}, \citenamefont {Rakhmilevitch},
  \citenamefont {Palevski},\ and\ \citenamefont
  {Dagan}}]{ben_shalom_tuning_2010}%
  \BibitemOpen
  \bibfield  {author} {\bibinfo {author} {\bibfnamefont {M.}~\bibnamefont
  {Ben~Shalom}}, \bibinfo {author} {\bibfnamefont {M.}~\bibnamefont {Sachs}},
  \bibinfo {author} {\bibfnamefont {D.}~\bibnamefont {Rakhmilevitch}}, \bibinfo
  {author} {\bibfnamefont {A.}~\bibnamefont {Palevski}},\ and\ \bibinfo
  {author} {\bibfnamefont {Y.}~\bibnamefont {Dagan}},\ }\href
  {https://doi.org/10.1103/PhysRevLett.104.126802} {\bibfield  {journal}
  {\bibinfo  {journal} {Physical Review Letters}\ }\textbf {\bibinfo {volume}
  {104}},\ \bibinfo {pages} {126802} (\bibinfo {year} {2010})}\BibitemShut
  {NoStop}%
\bibitem [{\citenamefont {Lesne}\ \emph {et~al.}(2016)\citenamefont {Lesne},
  \citenamefont {Fu}, \citenamefont {Oyarzun}, \citenamefont {Rojas-Sánchez},
  \citenamefont {Vaz}, \citenamefont {Naganuma}, \citenamefont {Sicoli},
  \citenamefont {Attané}, \citenamefont {Jamet}, \citenamefont {Jacquet},
  \citenamefont {George}, \citenamefont {Barthélémy}, \citenamefont
  {Jaffrès}, \citenamefont {Fert}, \citenamefont {Bibes},\ and\ \citenamefont
  {Vila}}]{lesne_highly_2016}%
  \BibitemOpen
  \bibfield  {author} {\bibinfo {author} {\bibfnamefont {E.}~\bibnamefont
  {Lesne}}, \bibinfo {author} {\bibfnamefont {Y.}~\bibnamefont {Fu}}, \bibinfo
  {author} {\bibfnamefont {S.}~\bibnamefont {Oyarzun}}, \bibinfo {author}
  {\bibfnamefont {J.~C.}\ \bibnamefont {Rojas-Sánchez}}, \bibinfo {author}
  {\bibfnamefont {D.~C.}\ \bibnamefont {Vaz}}, \bibinfo {author} {\bibfnamefont
  {H.}~\bibnamefont {Naganuma}}, \bibinfo {author} {\bibfnamefont
  {G.}~\bibnamefont {Sicoli}}, \bibinfo {author} {\bibfnamefont {J.-P.}\
  \bibnamefont {Attané}}, \bibinfo {author} {\bibfnamefont {M.}~\bibnamefont
  {Jamet}}, \bibinfo {author} {\bibfnamefont {E.}~\bibnamefont {Jacquet}},
  \bibinfo {author} {\bibfnamefont {J.-M.}\ \bibnamefont {George}}, \bibinfo
  {author} {\bibfnamefont {A.}~\bibnamefont {Barthélémy}}, \bibinfo {author}
  {\bibfnamefont {H.}~\bibnamefont {Jaffrès}}, \bibinfo {author}
  {\bibfnamefont {A.}~\bibnamefont {Fert}}, \bibinfo {author} {\bibfnamefont
  {M.}~\bibnamefont {Bibes}},\ and\ \bibinfo {author} {\bibfnamefont
  {L.}~\bibnamefont {Vila}},\ }\href {https://doi.org/10.1038/nmat4726}
  {\bibfield  {journal} {\bibinfo  {journal} {Nature Materials}\ }\textbf
  {\bibinfo {volume} {15}},\ \bibinfo {pages} {1261} (\bibinfo {year}
  {2016})}\BibitemShut {NoStop}%
\bibitem [{\citenamefont {Song}\ \emph {et~al.}(2017)\citenamefont {Song},
  \citenamefont {Zhang}, \citenamefont {Su}, \citenamefont {Yuan},
  \citenamefont {Chen}, \citenamefont {Xing}, \citenamefont {Shi},
  \citenamefont {Sun},\ and\ \citenamefont {Han}}]{song_observation_2017}%
  \BibitemOpen
  \bibfield  {author} {\bibinfo {author} {\bibfnamefont {Q.}~\bibnamefont
  {Song}}, \bibinfo {author} {\bibfnamefont {H.}~\bibnamefont {Zhang}},
  \bibinfo {author} {\bibfnamefont {T.}~\bibnamefont {Su}}, \bibinfo {author}
  {\bibfnamefont {W.}~\bibnamefont {Yuan}}, \bibinfo {author} {\bibfnamefont
  {Y.}~\bibnamefont {Chen}}, \bibinfo {author} {\bibfnamefont {W.}~\bibnamefont
  {Xing}}, \bibinfo {author} {\bibfnamefont {J.}~\bibnamefont {Shi}}, \bibinfo
  {author} {\bibfnamefont {J.}~\bibnamefont {Sun}},\ and\ \bibinfo {author}
  {\bibfnamefont {W.}~\bibnamefont {Han}},\ }\href
  {https://doi.org/10.1126/sciadv.1602312} {\bibfield  {journal} {\bibinfo
  {journal} {Science Advances}\ }\textbf {\bibinfo {volume} {3}},\ \bibinfo
  {pages} {e1602312} (\bibinfo {year} {2017})}\BibitemShut {NoStop}%
\bibitem [{\citenamefont {Goswami}\ \emph {et~al.}(2016)\citenamefont
  {Goswami}, \citenamefont {Mulazimoglu}, \citenamefont {Monteiro},
  \citenamefont {Wölbing}, \citenamefont {Koelle}, \citenamefont {Kleiner},
  \citenamefont {Blanter}, \citenamefont {Vandersypen},\ and\ \citenamefont
  {Caviglia}}]{goswami_quantum_2016}%
  \BibitemOpen
  \bibfield  {author} {\bibinfo {author} {\bibfnamefont {S.}~\bibnamefont
  {Goswami}}, \bibinfo {author} {\bibfnamefont {E.}~\bibnamefont
  {Mulazimoglu}}, \bibinfo {author} {\bibfnamefont {A.~M. R. V.~L.}\
  \bibnamefont {Monteiro}}, \bibinfo {author} {\bibfnamefont {R.}~\bibnamefont
  {Wölbing}}, \bibinfo {author} {\bibfnamefont {D.}~\bibnamefont {Koelle}},
  \bibinfo {author} {\bibfnamefont {R.}~\bibnamefont {Kleiner}}, \bibinfo
  {author} {\bibfnamefont {Y.~M.}\ \bibnamefont {Blanter}}, \bibinfo {author}
  {\bibfnamefont {L.~M.~K.}\ \bibnamefont {Vandersypen}},\ and\ \bibinfo
  {author} {\bibfnamefont {A.~D.}\ \bibnamefont {Caviglia}},\ }\href
  {https://doi.org/10.1038/nnano.2016.112} {\bibfield  {journal} {\bibinfo
  {journal} {Nature Nanotechnology}\ }\textbf {\bibinfo {volume} {11}},\
  \bibinfo {pages} {861} (\bibinfo {year} {2016})}\BibitemShut {NoStop}%
\bibitem [{\citenamefont {Brinkman}\ \emph {et~al.}(2007)\citenamefont
  {Brinkman}, \citenamefont {Huijben}, \citenamefont {van Zalk}, \citenamefont
  {Huijben}, \citenamefont {Zeitler}, \citenamefont {Maan}, \citenamefont
  {van~der Wiel}, \citenamefont {Rijnders}, \citenamefont {Blank},\ and\
  \citenamefont {Hilgenkamp}}]{brinkman_magnetic_2007}%
  \BibitemOpen
  \bibfield  {author} {\bibinfo {author} {\bibfnamefont {A.}~\bibnamefont
  {Brinkman}}, \bibinfo {author} {\bibfnamefont {M.}~\bibnamefont {Huijben}},
  \bibinfo {author} {\bibfnamefont {M.}~\bibnamefont {van Zalk}}, \bibinfo
  {author} {\bibfnamefont {J.}~\bibnamefont {Huijben}}, \bibinfo {author}
  {\bibfnamefont {U.}~\bibnamefont {Zeitler}}, \bibinfo {author} {\bibfnamefont
  {J.~C.}\ \bibnamefont {Maan}}, \bibinfo {author} {\bibfnamefont {W.~G.}\
  \bibnamefont {van~der Wiel}}, \bibinfo {author} {\bibfnamefont
  {G.}~\bibnamefont {Rijnders}}, \bibinfo {author} {\bibfnamefont {D.~H.~A.}\
  \bibnamefont {Blank}},\ and\ \bibinfo {author} {\bibfnamefont
  {H.}~\bibnamefont {Hilgenkamp}},\ }\href {https://doi.org/10.1038/nmat1931}
  {\bibfield  {journal} {\bibinfo  {journal} {Nature Materials}\ }\textbf
  {\bibinfo {volume} {6}},\ \bibinfo {pages} {493} (\bibinfo {year}
  {2007})}\BibitemShut {NoStop}%
\bibitem [{\citenamefont {Dikin}\ \emph {et~al.}(2011)\citenamefont {Dikin},
  \citenamefont {Mehta}, \citenamefont {Bark}, \citenamefont {Folkman},
  \citenamefont {Eom},\ and\ \citenamefont
  {Chandrasekhar}}]{dikin_coexistence_2011}%
  \BibitemOpen
  \bibfield  {author} {\bibinfo {author} {\bibfnamefont {D.~A.}\ \bibnamefont
  {Dikin}}, \bibinfo {author} {\bibfnamefont {M.}~\bibnamefont {Mehta}},
  \bibinfo {author} {\bibfnamefont {C.~W.}\ \bibnamefont {Bark}}, \bibinfo
  {author} {\bibfnamefont {C.~M.}\ \bibnamefont {Folkman}}, \bibinfo {author}
  {\bibfnamefont {C.~B.}\ \bibnamefont {Eom}},\ and\ \bibinfo {author}
  {\bibfnamefont {V.}~\bibnamefont {Chandrasekhar}},\ }\href
  {https://doi.org/10.1103/PhysRevLett.107.056802} {\bibfield  {journal}
  {\bibinfo  {journal} {Physical Review Letters}\ }\textbf {\bibinfo {volume}
  {107}},\ \bibinfo {pages} {056802} (\bibinfo {year} {2011})}\BibitemShut
  {NoStop}%
\bibitem [{\citenamefont {Bert}\ \emph {et~al.}(2011)\citenamefont {Bert},
  \citenamefont {Kalisky}, \citenamefont {Bell}, \citenamefont {Kim},
  \citenamefont {Hikita}, \citenamefont {Hwang},\ and\ \citenamefont
  {Moler}}]{bert_direct_2011}%
  \BibitemOpen
  \bibfield  {author} {\bibinfo {author} {\bibfnamefont {J.~A.}\ \bibnamefont
  {Bert}}, \bibinfo {author} {\bibfnamefont {B.}~\bibnamefont {Kalisky}},
  \bibinfo {author} {\bibfnamefont {C.}~\bibnamefont {Bell}}, \bibinfo {author}
  {\bibfnamefont {M.}~\bibnamefont {Kim}}, \bibinfo {author} {\bibfnamefont
  {Y.}~\bibnamefont {Hikita}}, \bibinfo {author} {\bibfnamefont {H.~Y.}\
  \bibnamefont {Hwang}},\ and\ \bibinfo {author} {\bibfnamefont {K.~A.}\
  \bibnamefont {Moler}},\ }\href {https://doi.org/10.1038/nphys2079} {\bibfield
   {journal} {\bibinfo  {journal} {Nature Physics}\ }\textbf {\bibinfo {volume}
  {7}},\ \bibinfo {pages} {767} (\bibinfo {year} {2011})}\BibitemShut {NoStop}%
\bibitem [{\citenamefont {Li}\ \emph {et~al.}(2011)\citenamefont {Li},
  \citenamefont {Richter}, \citenamefont {Mannhart},\ and\ \citenamefont
  {Ashoori}}]{li_coexistence_2011}%
  \BibitemOpen
  \bibfield  {author} {\bibinfo {author} {\bibfnamefont {L.}~\bibnamefont
  {Li}}, \bibinfo {author} {\bibfnamefont {C.}~\bibnamefont {Richter}},
  \bibinfo {author} {\bibfnamefont {J.}~\bibnamefont {Mannhart}},\ and\
  \bibinfo {author} {\bibfnamefont {R.~C.}\ \bibnamefont {Ashoori}},\ }\href
  {https://doi.org/10.1038/nphys2080} {\bibfield  {journal} {\bibinfo
  {journal} {Nature Physics}\ }\textbf {\bibinfo {volume} {7}},\ \bibinfo
  {pages} {762} (\bibinfo {year} {2011})}\BibitemShut {NoStop}%
\bibitem [{\citenamefont {Zou}\ \emph {et~al.}(2015)\citenamefont {Zou},
  \citenamefont {Ismail-Beigi}, \citenamefont {Kisslinger}, \citenamefont
  {Shen}, \citenamefont {Su}, \citenamefont {Walker},\ and\ \citenamefont
  {Ahn}}]{zou_latio_2015}%
  \BibitemOpen
  \bibfield  {author} {\bibinfo {author} {\bibfnamefont {K.}~\bibnamefont
  {Zou}}, \bibinfo {author} {\bibfnamefont {S.}~\bibnamefont {Ismail-Beigi}},
  \bibinfo {author} {\bibfnamefont {K.}~\bibnamefont {Kisslinger}}, \bibinfo
  {author} {\bibfnamefont {X.}~\bibnamefont {Shen}}, \bibinfo {author}
  {\bibfnamefont {D.}~\bibnamefont {Su}}, \bibinfo {author} {\bibfnamefont
  {F.~J.}\ \bibnamefont {Walker}},\ and\ \bibinfo {author} {\bibfnamefont
  {C.~H.}\ \bibnamefont {Ahn}},\ }\href {https://doi.org/10.1063/1.4914310}
  {\bibfield  {journal} {\bibinfo  {journal} {APL Materials}\ }\textbf
  {\bibinfo {volume} {3}},\ \bibinfo {pages} {036104} (\bibinfo {year}
  {2015})}\BibitemShut {NoStop}%
\bibitem [{\citenamefont {Bareille}\ \emph {et~al.}(2015)\citenamefont
  {Bareille}, \citenamefont {Fortuna}, \citenamefont {Rödel}, \citenamefont
  {Bertran}, \citenamefont {Gabay}, \citenamefont {Cubelos}, \citenamefont
  {Taleb-Ibrahimi}, \citenamefont {Le~Fèvre}, \citenamefont {Bibes},
  \citenamefont {Barthélémy}, \citenamefont {Maroutian}, \citenamefont
  {Lecoeur}, \citenamefont {Rozenberg},\ and\ \citenamefont
  {Santander-Syro}}]{bareille_two-dimensional_2015}%
  \BibitemOpen
  \bibfield  {author} {\bibinfo {author} {\bibfnamefont {C.}~\bibnamefont
  {Bareille}}, \bibinfo {author} {\bibfnamefont {F.}~\bibnamefont {Fortuna}},
  \bibinfo {author} {\bibfnamefont {T.~C.}\ \bibnamefont {Rödel}}, \bibinfo
  {author} {\bibfnamefont {F.}~\bibnamefont {Bertran}}, \bibinfo {author}
  {\bibfnamefont {M.}~\bibnamefont {Gabay}}, \bibinfo {author} {\bibfnamefont
  {O.~H.}\ \bibnamefont {Cubelos}}, \bibinfo {author} {\bibfnamefont
  {A.}~\bibnamefont {Taleb-Ibrahimi}}, \bibinfo {author} {\bibfnamefont
  {P.}~\bibnamefont {Le~Fèvre}}, \bibinfo {author} {\bibfnamefont
  {M.}~\bibnamefont {Bibes}}, \bibinfo {author} {\bibfnamefont
  {A.}~\bibnamefont {Barthélémy}}, \bibinfo {author} {\bibfnamefont
  {T.}~\bibnamefont {Maroutian}}, \bibinfo {author} {\bibfnamefont
  {P.}~\bibnamefont {Lecoeur}}, \bibinfo {author} {\bibfnamefont {M.~J.}\
  \bibnamefont {Rozenberg}},\ and\ \bibinfo {author} {\bibfnamefont {A.~F.}\
  \bibnamefont {Santander-Syro}},\ }\href {https://doi.org/10.1038/srep03586}
  {\bibfield  {journal} {\bibinfo  {journal} {Scientific Reports}\ }\textbf
  {\bibinfo {volume} {4}},\ \bibinfo {pages} {3586} (\bibinfo {year}
  {2015})}\BibitemShut {NoStop}%
\bibitem [{\citenamefont {Wadehra}\ and\ \citenamefont
  {Chakraverty}(2021)}]{wadehra_emergent_2021}%
  \BibitemOpen
  \bibfield  {author} {\bibinfo {author} {\bibfnamefont {N.}~\bibnamefont
  {Wadehra}}\ and\ \bibinfo {author} {\bibfnamefont {S.}~\bibnamefont
  {Chakraverty}},\ }\href {https://doi.org/10.1007/s12034-021-02564-6}
  {\bibfield  {journal} {\bibinfo  {journal} {Bulletin of Materials Science}\
  }\textbf {\bibinfo {volume} {44}},\ \bibinfo {pages} {269} (\bibinfo {year}
  {2021})}\BibitemShut {NoStop}%
\bibitem [{\citenamefont {Gupta}\ \emph {et~al.}(2022)\citenamefont {Gupta},
  \citenamefont {Silotia}, \citenamefont {Kumari}, \citenamefont {Dumen},
  \citenamefont {Goyal}, \citenamefont {Tomar}, \citenamefont {Wadehra},
  \citenamefont {Ayyub},\ and\ \citenamefont {Chakraverty}}]{gupta_ktao_2022}%
  \BibitemOpen
  \bibfield  {author} {\bibinfo {author} {\bibfnamefont {A.}~\bibnamefont
  {Gupta}}, \bibinfo {author} {\bibfnamefont {H.}~\bibnamefont {Silotia}},
  \bibinfo {author} {\bibfnamefont {A.}~\bibnamefont {Kumari}}, \bibinfo
  {author} {\bibfnamefont {M.}~\bibnamefont {Dumen}}, \bibinfo {author}
  {\bibfnamefont {S.}~\bibnamefont {Goyal}}, \bibinfo {author} {\bibfnamefont
  {R.}~\bibnamefont {Tomar}}, \bibinfo {author} {\bibfnamefont
  {N.}~\bibnamefont {Wadehra}}, \bibinfo {author} {\bibfnamefont
  {P.}~\bibnamefont {Ayyub}},\ and\ \bibinfo {author} {\bibfnamefont
  {S.}~\bibnamefont {Chakraverty}},\ }\href
  {https://doi.org/10.1002/adma.202106481} {\bibfield  {journal} {\bibinfo
  {journal} {Advanced Materials}\ }\textbf {\bibinfo {volume} {34}},\ \bibinfo
  {pages} {2106481} (\bibinfo {year} {2022})}\BibitemShut {NoStop}%
\bibitem [{\citenamefont {Wemple}(1965)}]{wemple_transport_1965}%
  \BibitemOpen
  \bibfield  {author} {\bibinfo {author} {\bibfnamefont {S.~H.}\ \bibnamefont
  {Wemple}},\ }\href {https://doi.org/10.1103/PhysRev.137.A1575} {\bibfield
  {journal} {\bibinfo  {journal} {Physical Review}\ }\textbf {\bibinfo {volume}
  {137}},\ \bibinfo {pages} {A1575} (\bibinfo {year} {1965})}\BibitemShut
  {NoStop}%
\bibitem [{\citenamefont {Fujii}\ and\ \citenamefont
  {Sakudo}(1976)}]{fujii_KTO_1976}%
  \BibitemOpen
  \bibfield  {author} {\bibinfo {author} {\bibfnamefont {Y.}~\bibnamefont
  {Fujii}}\ and\ \bibinfo {author} {\bibfnamefont {T.}~\bibnamefont {Sakudo}},\
  }\href {https://doi.org/10.1143/JPSJ.41.888} {\bibfield  {journal} {\bibinfo
  {journal} {Journal of the Physical Society of Japan}\ }\textbf {\bibinfo
  {volume} {41}},\ \bibinfo {pages} {888} (\bibinfo {year} {1976})},\ \Eprint
  {https://arxiv.org/abs/https://doi.org/10.1143/JPSJ.41.888}
  {https://doi.org/10.1143/JPSJ.41.888} \BibitemShut {NoStop}%
\bibitem [{\citenamefont {Fujishita}\ \emph {et~al.}(2016)\citenamefont
  {Fujishita}, \citenamefont {Kitazawa}, \citenamefont {Saito}, \citenamefont
  {Ishisaka}, \citenamefont {Okamoto},\ and\ \citenamefont
  {Yamaguchi}}]{fujishita_2016}%
  \BibitemOpen
  \bibfield  {author} {\bibinfo {author} {\bibfnamefont {H.}~\bibnamefont
  {Fujishita}}, \bibinfo {author} {\bibfnamefont {S.}~\bibnamefont {Kitazawa}},
  \bibinfo {author} {\bibfnamefont {M.}~\bibnamefont {Saito}}, \bibinfo
  {author} {\bibfnamefont {R.}~\bibnamefont {Ishisaka}}, \bibinfo {author}
  {\bibfnamefont {H.}~\bibnamefont {Okamoto}},\ and\ \bibinfo {author}
  {\bibfnamefont {T.}~\bibnamefont {Yamaguchi}},\ }\href
  {https://doi.org/10.7566/JPSJ.85.074703} {\bibfield  {journal} {\bibinfo
  {journal} {Journal of the Physical Society of Japan}\ }\textbf {\bibinfo
  {volume} {85}},\ \bibinfo {pages} {074703} (\bibinfo {year} {2016})},\
  \Eprint {https://arxiv.org/abs/https://doi.org/10.7566/JPSJ.85.074703}
  {https://doi.org/10.7566/JPSJ.85.074703} \BibitemShut {NoStop}%
\bibitem [{\citenamefont {Liu}\ \emph {et~al.}(2021)\citenamefont {Liu},
  \citenamefont {Yan}, \citenamefont {Jin}, \citenamefont {Ma}, \citenamefont
  {Hsiao}, \citenamefont {Lin}, \citenamefont {Bretz-Sullivan}, \citenamefont
  {Zhou}, \citenamefont {Pearson}, \citenamefont {Fisher}, \citenamefont
  {Jiang}, \citenamefont {Han}, \citenamefont {Zuo}, \citenamefont {Wen},
  \citenamefont {Fong}, \citenamefont {Sun}, \citenamefont {Zhou},\ and\
  \citenamefont {Bhattacharya}}]{liu_two_dimensional_2021}%
  \BibitemOpen
  \bibfield  {author} {\bibinfo {author} {\bibfnamefont {C.}~\bibnamefont
  {Liu}}, \bibinfo {author} {\bibfnamefont {X.}~\bibnamefont {Yan}}, \bibinfo
  {author} {\bibfnamefont {D.}~\bibnamefont {Jin}}, \bibinfo {author}
  {\bibfnamefont {Y.}~\bibnamefont {Ma}}, \bibinfo {author} {\bibfnamefont
  {H.-W.}\ \bibnamefont {Hsiao}}, \bibinfo {author} {\bibfnamefont
  {Y.}~\bibnamefont {Lin}}, \bibinfo {author} {\bibfnamefont {T.~M.}\
  \bibnamefont {Bretz-Sullivan}}, \bibinfo {author} {\bibfnamefont
  {X.}~\bibnamefont {Zhou}}, \bibinfo {author} {\bibfnamefont {J.}~\bibnamefont
  {Pearson}}, \bibinfo {author} {\bibfnamefont {B.}~\bibnamefont {Fisher}},
  \bibinfo {author} {\bibfnamefont {J.~S.}\ \bibnamefont {Jiang}}, \bibinfo
  {author} {\bibfnamefont {W.}~\bibnamefont {Han}}, \bibinfo {author}
  {\bibfnamefont {J.-M.}\ \bibnamefont {Zuo}}, \bibinfo {author} {\bibfnamefont
  {J.}~\bibnamefont {Wen}}, \bibinfo {author} {\bibfnamefont {D.~D.}\
  \bibnamefont {Fong}}, \bibinfo {author} {\bibfnamefont {J.}~\bibnamefont
  {Sun}}, \bibinfo {author} {\bibfnamefont {H.}~\bibnamefont {Zhou}},\ and\
  \bibinfo {author} {\bibfnamefont {A.}~\bibnamefont {Bhattacharya}},\ }\href
  {https://doi.org/10.1126/science.aba5511} {\bibfield  {journal} {\bibinfo
  {journal} {Science}\ }\textbf {\bibinfo {volume} {371}},\ \bibinfo {pages}
  {716} (\bibinfo {year} {2021})}\BibitemShut {NoStop}%
\bibitem [{\citenamefont {Zhang}\ \emph {et~al.}(2018)\citenamefont {Zhang},
  \citenamefont {Yun}, \citenamefont {Zhang}, \citenamefont {Zhang},
  \citenamefont {Ma}, \citenamefont {Yan}, \citenamefont {Wang}, \citenamefont
  {Li}, \citenamefont {Li}, \citenamefont {Khan}, \citenamefont {Chen},
  \citenamefont {Liu}, \citenamefont {Hu}, \citenamefont {Liu}, \citenamefont
  {Shen}, \citenamefont {Han},\ and\ \citenamefont
  {Sun}}]{zhang_high_mobility_2018}%
  \BibitemOpen
  \bibfield  {author} {\bibinfo {author} {\bibfnamefont {H.}~\bibnamefont
  {Zhang}}, \bibinfo {author} {\bibfnamefont {Y.}~\bibnamefont {Yun}}, \bibinfo
  {author} {\bibfnamefont {X.}~\bibnamefont {Zhang}}, \bibinfo {author}
  {\bibfnamefont {H.}~\bibnamefont {Zhang}}, \bibinfo {author} {\bibfnamefont
  {Y.}~\bibnamefont {Ma}}, \bibinfo {author} {\bibfnamefont {X.}~\bibnamefont
  {Yan}}, \bibinfo {author} {\bibfnamefont {F.}~\bibnamefont {Wang}}, \bibinfo
  {author} {\bibfnamefont {G.}~\bibnamefont {Li}}, \bibinfo {author}
  {\bibfnamefont {R.}~\bibnamefont {Li}}, \bibinfo {author} {\bibfnamefont
  {T.}~\bibnamefont {Khan}}, \bibinfo {author} {\bibfnamefont {Y.}~\bibnamefont
  {Chen}}, \bibinfo {author} {\bibfnamefont {W.}~\bibnamefont {Liu}}, \bibinfo
  {author} {\bibfnamefont {F.}~\bibnamefont {Hu}}, \bibinfo {author}
  {\bibfnamefont {B.}~\bibnamefont {Liu}}, \bibinfo {author} {\bibfnamefont
  {B.}~\bibnamefont {Shen}}, \bibinfo {author} {\bibfnamefont {W.}~\bibnamefont
  {Han}},\ and\ \bibinfo {author} {\bibfnamefont {J.}~\bibnamefont {Sun}},\
  }\href {https://doi.org/10.1103/PhysRevLett.121.116803} {\bibfield  {journal}
  {\bibinfo  {journal} {Physical Review Letters}\ }\textbf {\bibinfo {volume}
  {121}},\ \bibinfo {pages} {116803} (\bibinfo {year} {2018})}\BibitemShut
  {NoStop}%
\bibitem [{\citenamefont {Bruno}\ \emph {et~al.}(2019)\citenamefont {Bruno},
  \citenamefont {McKeown~Walker}, \citenamefont {Riccò}, \citenamefont
  {la~Torre}, \citenamefont {Wang}, \citenamefont {Tamai}, \citenamefont {Kim},
  \citenamefont {Hoesch}, \citenamefont {Bahramy},\ and\ \citenamefont
  {Baumberger}}]{bruno_band_2019}%
  \BibitemOpen
  \bibfield  {author} {\bibinfo {author} {\bibfnamefont {F.~Y.}\ \bibnamefont
  {Bruno}}, \bibinfo {author} {\bibfnamefont {S.}~\bibnamefont
  {McKeown~Walker}}, \bibinfo {author} {\bibfnamefont {S.}~\bibnamefont
  {Riccò}}, \bibinfo {author} {\bibfnamefont {A.}~\bibnamefont {la~Torre}},
  \bibinfo {author} {\bibfnamefont {Z.}~\bibnamefont {Wang}}, \bibinfo {author}
  {\bibfnamefont {A.}~\bibnamefont {Tamai}}, \bibinfo {author} {\bibfnamefont
  {T.~K.}\ \bibnamefont {Kim}}, \bibinfo {author} {\bibfnamefont
  {M.}~\bibnamefont {Hoesch}}, \bibinfo {author} {\bibfnamefont {M.~S.}\
  \bibnamefont {Bahramy}},\ and\ \bibinfo {author} {\bibfnamefont
  {F.}~\bibnamefont {Baumberger}},\ }\href
  {https://doi.org/10.1002/aelm.201800860} {\bibfield  {journal} {\bibinfo
  {journal} {Advanced Electronic Materials}\ }\textbf {\bibinfo {volume} {5}},\
  \bibinfo {pages} {1800860} (\bibinfo {year} {2019})}\BibitemShut {NoStop}%
\bibitem [{\citenamefont {Nakamura}\ and\ \citenamefont
  {Kimura}(2009)}]{nakamura_electric_2009}%
  \BibitemOpen
  \bibfield  {author} {\bibinfo {author} {\bibfnamefont {H.}~\bibnamefont
  {Nakamura}}\ and\ \bibinfo {author} {\bibfnamefont {T.}~\bibnamefont
  {Kimura}},\ }\href {https://doi.org/10.1103/PhysRevB.80.121308} {\bibfield
  {journal} {\bibinfo  {journal} {Physical Review B}\ }\textbf {\bibinfo
  {volume} {80}},\ \bibinfo {pages} {121308} (\bibinfo {year}
  {2009})}\BibitemShut {NoStop}%
\bibitem [{\citenamefont {Davis}\ \emph {et~al.}(2017)\citenamefont {Davis},
  \citenamefont {Chandrasekhar}, \citenamefont {Huang}, \citenamefont {Han},
  \citenamefont {{Ariando}},\ and\ \citenamefont
  {Venkatesan}}]{davis_anisotropic_2017}%
  \BibitemOpen
  \bibfield  {author} {\bibinfo {author} {\bibfnamefont {S.}~\bibnamefont
  {Davis}}, \bibinfo {author} {\bibfnamefont {V.}~\bibnamefont
  {Chandrasekhar}}, \bibinfo {author} {\bibfnamefont {Z.}~\bibnamefont
  {Huang}}, \bibinfo {author} {\bibfnamefont {K.}~\bibnamefont {Han}}, \bibinfo
  {author} {\bibnamefont {{Ariando}}},\ and\ \bibinfo {author} {\bibfnamefont
  {T.}~\bibnamefont {Venkatesan}},\ }\href
  {https://doi.org/10.1103/PhysRevB.95.035127} {\bibfield  {journal} {\bibinfo
  {journal} {Physical Review B}\ }\textbf {\bibinfo {volume} {95}},\ \bibinfo
  {pages} {035127} (\bibinfo {year} {2017})}\BibitemShut {NoStop}%
\bibitem [{\citenamefont {Krantz}\ and\ \citenamefont
  {Chandrasekhar}(2021)}]{krantz_observation_2021}%
  \BibitemOpen
  \bibfield  {author} {\bibinfo {author} {\bibfnamefont {P.}~\bibnamefont
  {Krantz}}\ and\ \bibinfo {author} {\bibfnamefont {V.}~\bibnamefont
  {Chandrasekhar}},\ }\href {https://doi.org/10.1103/PhysRevLett.127.036801}
  {\bibfield  {journal} {\bibinfo  {journal} {Physical Review Letters}\
  }\textbf {\bibinfo {volume} {127}},\ \bibinfo {pages} {036801} (\bibinfo
  {year} {2021})}\BibitemShut {NoStop}%
\bibitem [{\citenamefont {Biscaras}\ \emph {et~al.}(2015)\citenamefont
  {Biscaras}, \citenamefont {Hurand}, \citenamefont {Feuillet-Palma},
  \citenamefont {Rastogi}, \citenamefont {Budhani}, \citenamefont {Reyren},
  \citenamefont {Lesne}, \citenamefont {Lesueur},\ and\ \citenamefont
  {Bergeal}}]{biscaras_limit_2015}%
  \BibitemOpen
  \bibfield  {author} {\bibinfo {author} {\bibfnamefont {J.}~\bibnamefont
  {Biscaras}}, \bibinfo {author} {\bibfnamefont {S.}~\bibnamefont {Hurand}},
  \bibinfo {author} {\bibfnamefont {C.}~\bibnamefont {Feuillet-Palma}},
  \bibinfo {author} {\bibfnamefont {A.}~\bibnamefont {Rastogi}}, \bibinfo
  {author} {\bibfnamefont {R.~C.}\ \bibnamefont {Budhani}}, \bibinfo {author}
  {\bibfnamefont {N.}~\bibnamefont {Reyren}}, \bibinfo {author} {\bibfnamefont
  {E.}~\bibnamefont {Lesne}}, \bibinfo {author} {\bibfnamefont
  {J.}~\bibnamefont {Lesueur}},\ and\ \bibinfo {author} {\bibfnamefont
  {N.}~\bibnamefont {Bergeal}},\ }\href {https://doi.org/10.1038/srep06788}
  {\bibfield  {journal} {\bibinfo  {journal} {Scientific Reports}\ }\textbf
  {\bibinfo {volume} {4}},\ \bibinfo {pages} {6788} (\bibinfo {year}
  {2015})}\BibitemShut {NoStop}%
\bibitem [{\citenamefont {Qiao}\ \emph {et~al.}(2021)\citenamefont {Qiao},
  \citenamefont {Ma}, \citenamefont {Yan}, \citenamefont {Xing}, \citenamefont
  {Yao}, \citenamefont {Cai}, \citenamefont {Li}, \citenamefont {Xiong},
  \citenamefont {Xie}, \citenamefont {Lin},\ and\ \citenamefont
  {Han}}]{qiao_gate_2021}%
  \BibitemOpen
  \bibfield  {author} {\bibinfo {author} {\bibfnamefont {W.}~\bibnamefont
  {Qiao}}, \bibinfo {author} {\bibfnamefont {Y.}~\bibnamefont {Ma}}, \bibinfo
  {author} {\bibfnamefont {J.}~\bibnamefont {Yan}}, \bibinfo {author}
  {\bibfnamefont {W.}~\bibnamefont {Xing}}, \bibinfo {author} {\bibfnamefont
  {Y.}~\bibnamefont {Yao}}, \bibinfo {author} {\bibfnamefont {R.}~\bibnamefont
  {Cai}}, \bibinfo {author} {\bibfnamefont {B.}~\bibnamefont {Li}}, \bibinfo
  {author} {\bibfnamefont {R.}~\bibnamefont {Xiong}}, \bibinfo {author}
  {\bibfnamefont {X.~C.}\ \bibnamefont {Xie}}, \bibinfo {author} {\bibfnamefont
  {X.}~\bibnamefont {Lin}},\ and\ \bibinfo {author} {\bibfnamefont
  {W.}~\bibnamefont {Han}},\ }\href
  {https://doi.org/10.1103/PhysRevB.104.184505} {\bibfield  {journal} {\bibinfo
   {journal} {Physical Review B}\ }\textbf {\bibinfo {volume} {104}},\ \bibinfo
  {pages} {184505} (\bibinfo {year} {2021})}\BibitemShut {NoStop}%
\bibitem [{\citenamefont {Pavlenko}\ \emph {et~al.}(2012)\citenamefont
  {Pavlenko}, \citenamefont {Kopp}, \citenamefont {Tsymbal}, \citenamefont
  {Sawatzky},\ and\ \citenamefont {Mannhart}}]{pavlenko_magnetic_2012}%
  \BibitemOpen
  \bibfield  {author} {\bibinfo {author} {\bibfnamefont {N.}~\bibnamefont
  {Pavlenko}}, \bibinfo {author} {\bibfnamefont {T.}~\bibnamefont {Kopp}},
  \bibinfo {author} {\bibfnamefont {E.~Y.}\ \bibnamefont {Tsymbal}}, \bibinfo
  {author} {\bibfnamefont {G.~A.}\ \bibnamefont {Sawatzky}},\ and\ \bibinfo
  {author} {\bibfnamefont {J.}~\bibnamefont {Mannhart}},\ }\href
  {https://doi.org/10.1103/PhysRevB.85.020407} {\bibfield  {journal} {\bibinfo
  {journal} {Physical Review B}\ }\textbf {\bibinfo {volume} {85}},\ \bibinfo
  {pages} {020407} (\bibinfo {year} {2012})}\BibitemShut {NoStop}%
\bibitem [{\citenamefont {McGuire}\ and\ \citenamefont
  {Potter}(1975)}]{mcguire_anisotropic_1975}%
  \BibitemOpen
  \bibfield  {author} {\bibinfo {author} {\bibfnamefont {T.}~\bibnamefont
  {McGuire}}\ and\ \bibinfo {author} {\bibfnamefont {R.}~\bibnamefont
  {Potter}},\ }\href {https://doi.org/10.1109/TMAG.1975.1058782} {\bibfield
  {journal} {\bibinfo  {journal} {IEEE Transactions on Magnetics}\ }\textbf
  {\bibinfo {volume} {11}},\ \bibinfo {pages} {1018} (\bibinfo {year}
  {1975})}\BibitemShut {NoStop}%
\bibitem [{\citenamefont {Nagaosa}\ \emph {et~al.}(2010)\citenamefont
  {Nagaosa}, \citenamefont {Sinova}, \citenamefont {Onoda}, \citenamefont
  {MacDonald},\ and\ \citenamefont {Ong}}]{nagaosa_anomalous_2010}%
  \BibitemOpen
  \bibfield  {author} {\bibinfo {author} {\bibfnamefont {N.}~\bibnamefont
  {Nagaosa}}, \bibinfo {author} {\bibfnamefont {J.}~\bibnamefont {Sinova}},
  \bibinfo {author} {\bibfnamefont {S.}~\bibnamefont {Onoda}}, \bibinfo
  {author} {\bibfnamefont {A.~H.}\ \bibnamefont {MacDonald}},\ and\ \bibinfo
  {author} {\bibfnamefont {N.~P.}\ \bibnamefont {Ong}},\ }\href
  {https://doi.org/10.1103/RevModPhys.82.1539} {\bibfield  {journal} {\bibinfo
  {journal} {Reviews of Modern Physics}\ }\textbf {\bibinfo {volume} {82}},\
  \bibinfo {pages} {1539} (\bibinfo {year} {2010})}\BibitemShut {NoStop}%
\bibitem [{\citenamefont {Nakatsuji}\ \emph {et~al.}(2011)\citenamefont
  {Nakatsuji}, \citenamefont {Machida}, \citenamefont {Ishikawa}, \citenamefont
  {Onoda}, \citenamefont {Karaki}, \citenamefont {Tayama},\ and\ \citenamefont
  {Sakakibara}}]{nakatsuji_spontaneous_2011}%
  \BibitemOpen
  \bibfield  {author} {\bibinfo {author} {\bibfnamefont {S.}~\bibnamefont
  {Nakatsuji}}, \bibinfo {author} {\bibfnamefont {Y.}~\bibnamefont {Machida}},
  \bibinfo {author} {\bibfnamefont {J.~J.}\ \bibnamefont {Ishikawa}}, \bibinfo
  {author} {\bibfnamefont {S.}~\bibnamefont {Onoda}}, \bibinfo {author}
  {\bibfnamefont {Y.}~\bibnamefont {Karaki}}, \bibinfo {author} {\bibfnamefont
  {T.}~\bibnamefont {Tayama}},\ and\ \bibinfo {author} {\bibfnamefont
  {T.}~\bibnamefont {Sakakibara}},\ }\href
  {https://doi.org/10.1088/1742-6596/320/1/012056} {\bibfield  {journal}
  {\bibinfo  {journal} {Journal of Physics: Conference Series}\ }\textbf
  {\bibinfo {volume} {320}},\ \bibinfo {pages} {012056} (\bibinfo {year}
  {2011})}\BibitemShut {NoStop}%
\bibitem [{\citenamefont {Ueda}\ \emph {et~al.}(2018)\citenamefont {Ueda},
  \citenamefont {Kaneko}, \citenamefont {Ishizuka}, \citenamefont {Fujioka},
  \citenamefont {Nagaosa},\ and\ \citenamefont
  {Tokura}}]{ueda_spontaneous_2018}%
  \BibitemOpen
  \bibfield  {author} {\bibinfo {author} {\bibfnamefont {K.}~\bibnamefont
  {Ueda}}, \bibinfo {author} {\bibfnamefont {R.}~\bibnamefont {Kaneko}},
  \bibinfo {author} {\bibfnamefont {H.}~\bibnamefont {Ishizuka}}, \bibinfo
  {author} {\bibfnamefont {J.}~\bibnamefont {Fujioka}}, \bibinfo {author}
  {\bibfnamefont {N.}~\bibnamefont {Nagaosa}},\ and\ \bibinfo {author}
  {\bibfnamefont {Y.}~\bibnamefont {Tokura}},\ }\href
  {https://doi.org/10.1038/s41467-018-05530-9} {\bibfield  {journal} {\bibinfo
  {journal} {Nature Communications}\ }\textbf {\bibinfo {volume} {9}},\
  \bibinfo {pages} {3032} (\bibinfo {year} {2018})}\BibitemShut {NoStop}%
\bibitem [{\citenamefont {Dzsaber}\ \emph {et~al.}(2021)\citenamefont
  {Dzsaber}, \citenamefont {Yan}, \citenamefont {Taupin}, \citenamefont
  {Eguchi}, \citenamefont {Prokofiev}, \citenamefont {Shiroka}, \citenamefont
  {Blaha}, \citenamefont {Rubel}, \citenamefont {Grefe}, \citenamefont {Lai},
  \citenamefont {Si},\ and\ \citenamefont {Paschen}}]{dzsaber_giant_2021}%
  \BibitemOpen
  \bibfield  {author} {\bibinfo {author} {\bibfnamefont {S.}~\bibnamefont
  {Dzsaber}}, \bibinfo {author} {\bibfnamefont {X.}~\bibnamefont {Yan}},
  \bibinfo {author} {\bibfnamefont {M.}~\bibnamefont {Taupin}}, \bibinfo
  {author} {\bibfnamefont {G.}~\bibnamefont {Eguchi}}, \bibinfo {author}
  {\bibfnamefont {A.}~\bibnamefont {Prokofiev}}, \bibinfo {author}
  {\bibfnamefont {T.}~\bibnamefont {Shiroka}}, \bibinfo {author} {\bibfnamefont
  {P.}~\bibnamefont {Blaha}}, \bibinfo {author} {\bibfnamefont
  {O.}~\bibnamefont {Rubel}}, \bibinfo {author} {\bibfnamefont {S.~E.}\
  \bibnamefont {Grefe}}, \bibinfo {author} {\bibfnamefont {H.-H.}\ \bibnamefont
  {Lai}}, \bibinfo {author} {\bibfnamefont {Q.}~\bibnamefont {Si}},\ and\
  \bibinfo {author} {\bibfnamefont {S.}~\bibnamefont {Paschen}},\ }\href
  {https://doi.org/10.1073/pnas.2013386118} {\bibfield  {journal} {\bibinfo
  {journal} {Proceedings of the National Academy of Sciences}\ }\textbf
  {\bibinfo {volume} {118}},\ \bibinfo {pages} {e2013386118} (\bibinfo {year}
  {2021})}\BibitemShut {NoStop}%
\bibitem [{\citenamefont {Sodemann}\ and\ \citenamefont
  {Fu}(2015)}]{sodemann_quantum_2015}%
  \BibitemOpen
  \bibfield  {author} {\bibinfo {author} {\bibfnamefont {I.}~\bibnamefont
  {Sodemann}}\ and\ \bibinfo {author} {\bibfnamefont {L.}~\bibnamefont {Fu}},\
  }\href {https://doi.org/10.1103/PhysRevLett.115.216806} {\bibfield  {journal}
  {\bibinfo  {journal} {Physical Review Letters}\ }\textbf {\bibinfo {volume}
  {115}},\ \bibinfo {pages} {216806} (\bibinfo {year} {2015})}\BibitemShut
  {NoStop}%
\bibitem [{\citenamefont {Lesne}\ \emph {et~al.}(2022)\citenamefont {Lesne},
  \citenamefont {Sa\v{g}lam}, \citenamefont {Battilomo}, \citenamefont {van
  Thiel}, \citenamefont {Filippozzi}, \citenamefont {Cuoco}, \citenamefont
  {Steele}, \citenamefont {Ortix},\ and\ \citenamefont
  {Caviglia}}]{lesne_designing_2022}%
  \BibitemOpen
  \bibfield  {author} {\bibinfo {author} {\bibfnamefont {E.}~\bibnamefont
  {Lesne}}, \bibinfo {author} {\bibfnamefont {Y.~G.}\ \bibnamefont
  {Sa\v{g}lam}}, \bibinfo {author} {\bibfnamefont {R.}~\bibnamefont
  {Battilomo}}, \bibinfo {author} {\bibfnamefont {T.~C.}\ \bibnamefont {van
  Thiel}}, \bibinfo {author} {\bibfnamefont {U.}~\bibnamefont {Filippozzi}},
  \bibinfo {author} {\bibfnamefont {M.}~\bibnamefont {Cuoco}}, \bibinfo
  {author} {\bibfnamefont {G.~A.}\ \bibnamefont {Steele}}, \bibinfo {author}
  {\bibfnamefont {C.}~\bibnamefont {Ortix}},\ and\ \bibinfo {author}
  {\bibfnamefont {A.~D.}\ \bibnamefont {Caviglia}},\ }\href
  {http://arxiv.org/abs/2201.12161} {\bibinfo {title} {Designing {Berry}
  curvature dipoles and the quantum nonlinear {Hall} effect at oxide
  interfaces}} (\bibinfo {year} {2022}),\ \bibinfo {note} {arXiv:2201.12161
  [cond-mat]}\BibitemShut {NoStop}%
\bibitem [{\citenamefont {Ma}\ \emph {et~al.}(2019)\citenamefont {Ma},
  \citenamefont {Xu}, \citenamefont {Shen}, \citenamefont {MacNeill},
  \citenamefont {Fatemi}, \citenamefont {Chang}, \citenamefont {Mier~Valdivia},
  \citenamefont {Wu}, \citenamefont {Du}, \citenamefont {Hsu}, \citenamefont
  {Fang}, \citenamefont {Gibson}, \citenamefont {Watanabe}, \citenamefont
  {Taniguchi}, \citenamefont {Cava}, \citenamefont {Kaxiras}, \citenamefont
  {Lu}, \citenamefont {Lin}, \citenamefont {Fu}, \citenamefont {Gedik},\ and\
  \citenamefont {Jarillo-Herrero}}]{ma_observation_2019}%
  \BibitemOpen
  \bibfield  {author} {\bibinfo {author} {\bibfnamefont {Q.}~\bibnamefont
  {Ma}}, \bibinfo {author} {\bibfnamefont {S.-Y.}\ \bibnamefont {Xu}}, \bibinfo
  {author} {\bibfnamefont {H.}~\bibnamefont {Shen}}, \bibinfo {author}
  {\bibfnamefont {D.}~\bibnamefont {MacNeill}}, \bibinfo {author}
  {\bibfnamefont {V.}~\bibnamefont {Fatemi}}, \bibinfo {author} {\bibfnamefont
  {T.-R.}\ \bibnamefont {Chang}}, \bibinfo {author} {\bibfnamefont {A.~M.}\
  \bibnamefont {Mier~Valdivia}}, \bibinfo {author} {\bibfnamefont
  {S.}~\bibnamefont {Wu}}, \bibinfo {author} {\bibfnamefont {Z.}~\bibnamefont
  {Du}}, \bibinfo {author} {\bibfnamefont {C.-H.}\ \bibnamefont {Hsu}},
  \bibinfo {author} {\bibfnamefont {S.}~\bibnamefont {Fang}}, \bibinfo {author}
  {\bibfnamefont {Q.~D.}\ \bibnamefont {Gibson}}, \bibinfo {author}
  {\bibfnamefont {K.}~\bibnamefont {Watanabe}}, \bibinfo {author}
  {\bibfnamefont {T.}~\bibnamefont {Taniguchi}}, \bibinfo {author}
  {\bibfnamefont {R.~J.}\ \bibnamefont {Cava}}, \bibinfo {author}
  {\bibfnamefont {E.}~\bibnamefont {Kaxiras}}, \bibinfo {author} {\bibfnamefont
  {H.-Z.}\ \bibnamefont {Lu}}, \bibinfo {author} {\bibfnamefont
  {H.}~\bibnamefont {Lin}}, \bibinfo {author} {\bibfnamefont {L.}~\bibnamefont
  {Fu}}, \bibinfo {author} {\bibfnamefont {N.}~\bibnamefont {Gedik}},\ and\
  \bibinfo {author} {\bibfnamefont {P.}~\bibnamefont {Jarillo-Herrero}},\
  }\href {https://doi.org/10.1038/s41586-018-0807-6} {\bibfield  {journal}
  {\bibinfo  {journal} {Nature}\ }\textbf {\bibinfo {volume} {565}},\ \bibinfo
  {pages} {337} (\bibinfo {year} {2019})}\BibitemShut {NoStop}%
\end{thebibliography}%
\end{document}

\preprint{APS/123-QED}

\title{Supplemental Material: \\ Colossal Spontaneous Hall Effect and Emergent Magnetism in \\ KTaO$_3$ Two-Dimensional Electron Gases}

\author{Patrick W. Krantz}
\affiliation{Department of Physics, Northwestern University, Evanston, Illinois. 60208, USA
}%
\author{Alexander Tyner}%
\affiliation{Graduate Program for Applied Physics, Northwestern University, Evanston, Illinois. 60208, USA
}%
\author{Pallab Goswami}
\affiliation{Department of Physics, Northwestern University, Evanston, Illinois. 60208, USA
}%
\affiliation{Graduate Program for Applied Physics, Northwestern University, Evanston, Illinois. 60208, USA
}%
\author{Venkat Chandrasekhar}
\affiliation{Department of Physics, Northwestern University, Evanston, Illinois. 60208, USA
}%
\affiliation{Graduate Program for Applied Physics, Northwestern University, Evanston, Illinois. 60208, USA
}%

\date{\today}

\maketitle

\subsection*{Experimental Details}
\paragraph*{Sample Fabrication}
The KTaO$_3$ substrates used in this study were 5 mm x 5 mm x 0.5 mm single side polished crystals purchased from MSE Supplies LLC. For the samples described in the main text, the substrates were subjected to a normal cleaning regimen of 3 min ultrasonication in acetone, 3 min ultrasonication in DI water and 3 min ultrasonication in isopropanol before being spin coated with LOR-5A and S-1813 photoresists. The Hall bar patterns were exposed using a Suss MJB4 mask aligner before being developed in MF-319 developer. All three crystals were simultaneously metallized with 99.9995\% aluminum in an in-house deposition chamber via electron-gun evaporation. Before deposition the surface was cleaned with 100 mTorr of oxygen plasma for one minute to remove residue from the lithography process. Deposition consisted of repeated steps of 1.5 nm aluminum deposition followed by a dwell period of 10 min. After the second and subsequent steps a partial pressure of oxygen was introduce into the chamber to ensure oxidation of the aluminum layers to amorphous aluminum oxide. Samples thus defined were then all wirebonded to a sample header for simultaneous measurement and cooled down in an Oxford MX100 cryostat equipped with a two-axis superconducting magnet.
\newline{}

\paragraph*{Measurement Techniques}
Measurements were conducted using a low frequency AC lock-in technique with two PAR 124A lock-in amplifiers and an EGG 7260 digital lock-in amplifier. 100 nA of current was passed along the Hall bars sourced from home built current sources, and the resulting voltages were measured with similarly home built AD624 based pre-amplifier before being read by the lock-ins. The back gate voltage, $V_{g}$, was supplied by a Keithley KT2400 source equipped with a low pass filter and measured independently by a HP34401A digital multimeter.
\newline{}

\begin{figure}[h]
\centering
\includegraphics[width=0.55\columnwidth]{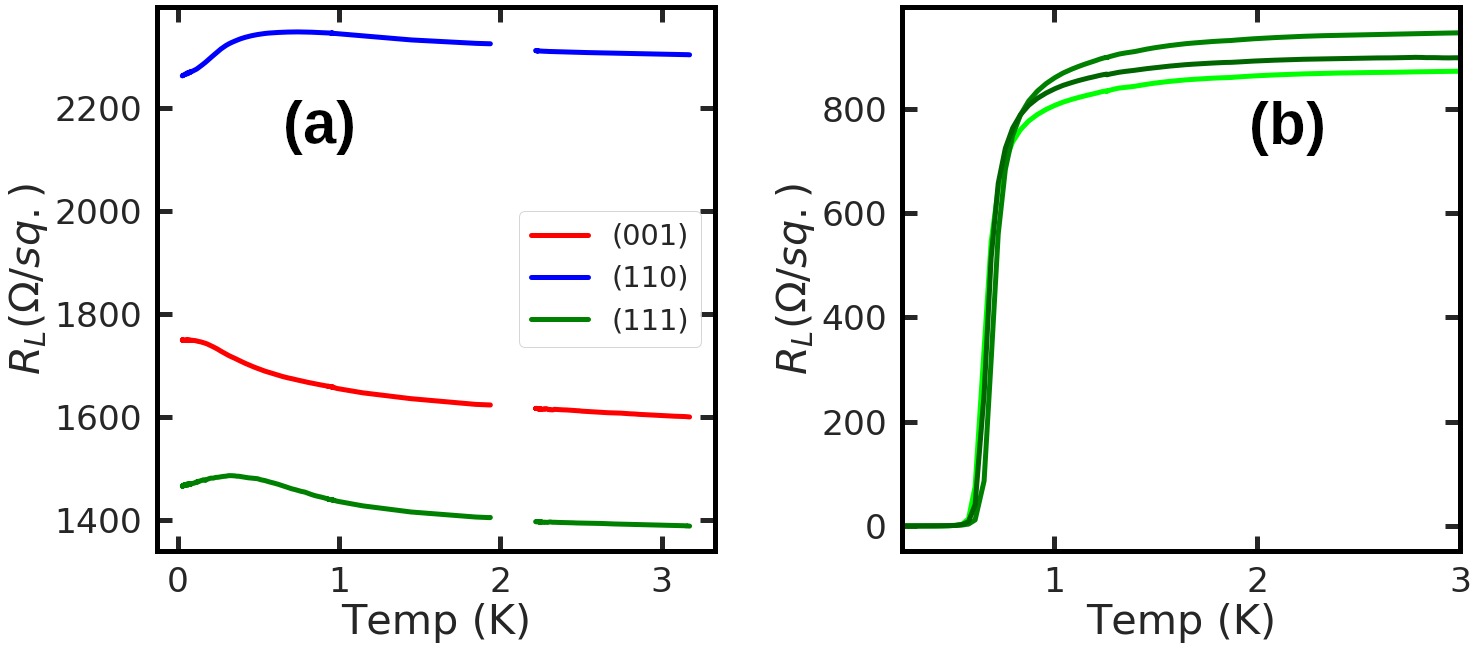}
\caption{\label{RvsT} (a) Sheet resistance of the three crystal terminations of the samples described in the main text after annealing sweeps of $V_g$. (b) Superconducting transition for a (111) sample fabricated with a 650C annealing step and deionized water soak to prepare the KTO surface.}
\end{figure}

\paragraph*{Temperature Dependence Resistance}
Resistance measurements as a function of temperature were taken for the reported samples after an initial gate voltage anneal cycle at $\sim$1.6K. These results are reported in Figure \ref{RvsT} (a), showing zero backgate voltage $R_L (T)$ for the two measured ranges from the cooldown steps on the dilution refrigerator. The resistance rises slightly from a minimum around 5K, before dipping or saturating at the lowest measured temperatures. Importantly, none of the three samples show superconducting transitions down to our lowest measurement temperature of 25mK.

\paragraph*{Superconducting Sample Preparation}
To demonstrate that the lack of annealing is the main factor in the suppression of the superconductivity for the samples described in the main text, additional samples were made with all of the same processes, but with an added annealing step before lithography. This additional annealing step consisted of a 650C bake for two hours in atmosphere before ultrasonication in DI water for two more hours as described by Tomar et al. [\textit{40}]. Subsequent fabrication steps were identical to those described above. The resulting samples were measured on a different cryostat, looking for a superconducting transition, which is shown for the (111) terminated samples in Figure \ref{RvsT} (b). The transition was measured on three different Hall bars all on the same (111) crystal substrate, but measured simultaneously. The data shown are for initial cooldown, before any annealing sweeps, with 100V backgate applied, as prescribed by other works on superconductivity in KTO.
\newline{}

\begin{figure}[h]
\centering
\includegraphics[width=0.30\columnwidth]{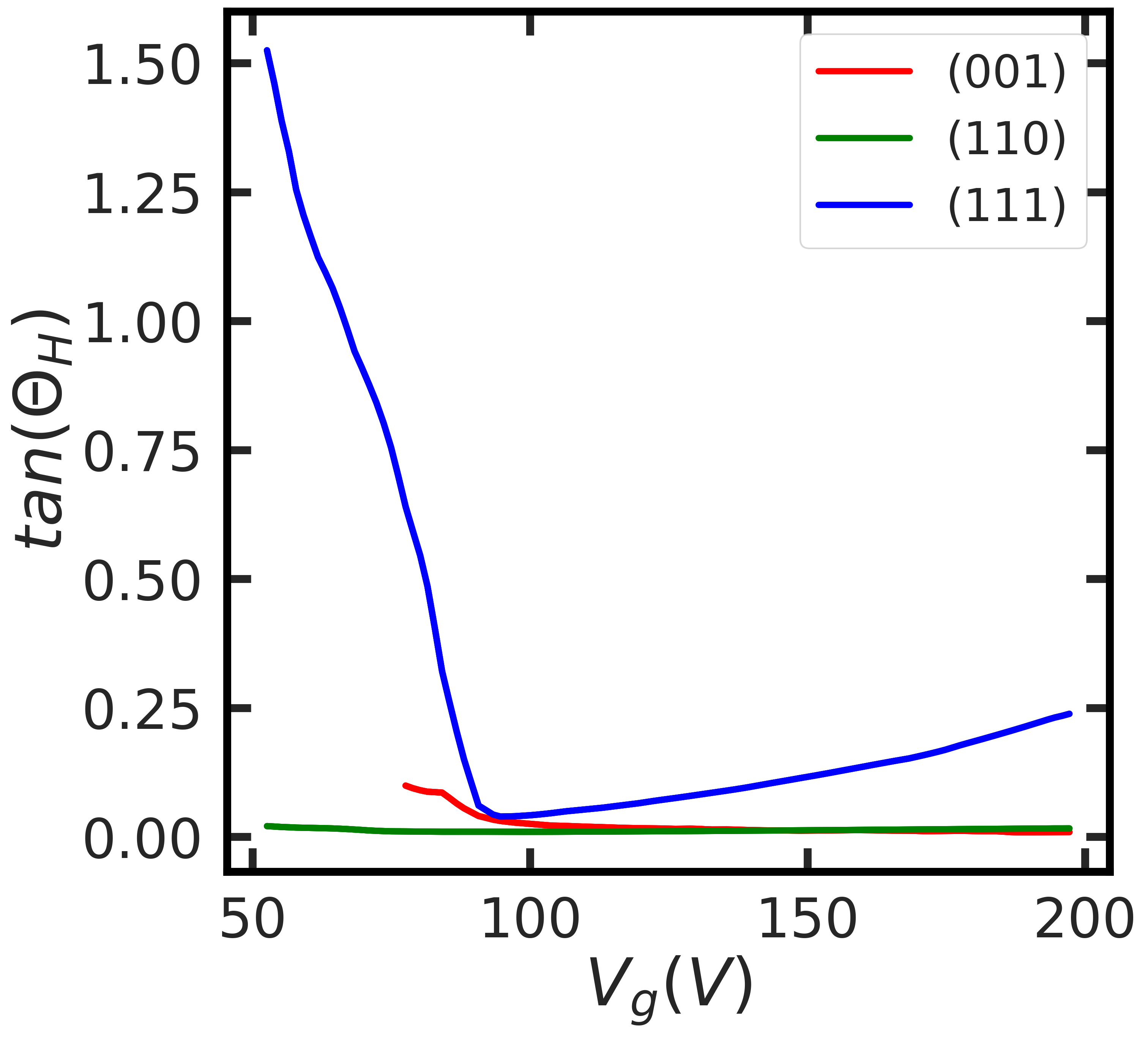}
\caption{\label{HallAngle} Calculated Hall Angle reported in $tan^{-1}(\rho_{xy}/\rho_{xx})$. Data was generated from those reported in Fig. 1 of the Main Text, for each of the (001), (110), and (111). The data have been averaged to remove the hysteresis, and the deviation behavior in the (001) below $\sim 70V$ was removed for clarity.}
\end{figure}

\paragraph*{Calculation of Hall Angle}
Hall angle is defined as $\Theta_H = \tan^{-1}(\sigma_{xy}/\sigma_{xx})$. In one dimension, $\sigma_{a} = \rho^{-1}_{a}$ but in two dimensions the conductivity \boldsymbol{$\sigma$} is a 2x2 matrix $\begin{pmatrix} 
  \sigma_{xx} & \sigma_{xy}\\ 
  \sigma_{yx} & \sigma_{yy}
\end{pmatrix}$, where Onsager’s reciprocity relations give $\sigma_{xx} = \sigma_{yy}$ and $\sigma_{xy} = \sigma_{yx}$ for an isotropic conductor. Similarly we can write the resistivity matrix \boldsymbol{$\rho$} as $\begin{pmatrix} 
  \rho_{xx} & \rho_{xy}\\ 
  -\rho_{yx} & \rho_{yy}
\end{pmatrix}$, and then inverting the \boldsymbol{$\rho$} matrix gives the conductivity matrix in terms of the measurable quantities $\rho_{xx}$ and $\rho_{xy}$ as $\frac{1}{\rho_{xx}\rho_{yy} - \rho^{2}_{xy}}\begin{pmatrix} 
  \rho_{xx} & \rho_{xy}\\ 
  -\rho_{yx} & \rho_{yy}
\end{pmatrix}$. It is then clear that we can replace $\Theta_H = \tan^{-1}(\sigma_{xy}/\sigma_{xx})$ with the directly measured $\Theta_H = \tan^{-1}(\rho_{xy}/\rho_{xx}) = \tan^{-1}(R_T/R_\square)$, which we can extract a function of gate voltage from the data shown in Fig. 1 of the Main Text ($R_\square$ is the longitudinal resistance per square). These results for the three crystal directions are reported in Fig. \ref{HallAngle}. We have removed the data from the (001) sample below $V_{g} \sim 80V$ due to the downturn of the longitudinal trace shown in Fig. 1 of the Main Text, which suggests an un-physical signal. Similar changes in the longitudinal are not seen in the (110) or the (111), even for larger values of resistance, so the source of the large Hall Angle is not similarly in question.

\begin{figure}[h]
	\centering
	\includegraphics[width=0.85\columnwidth]{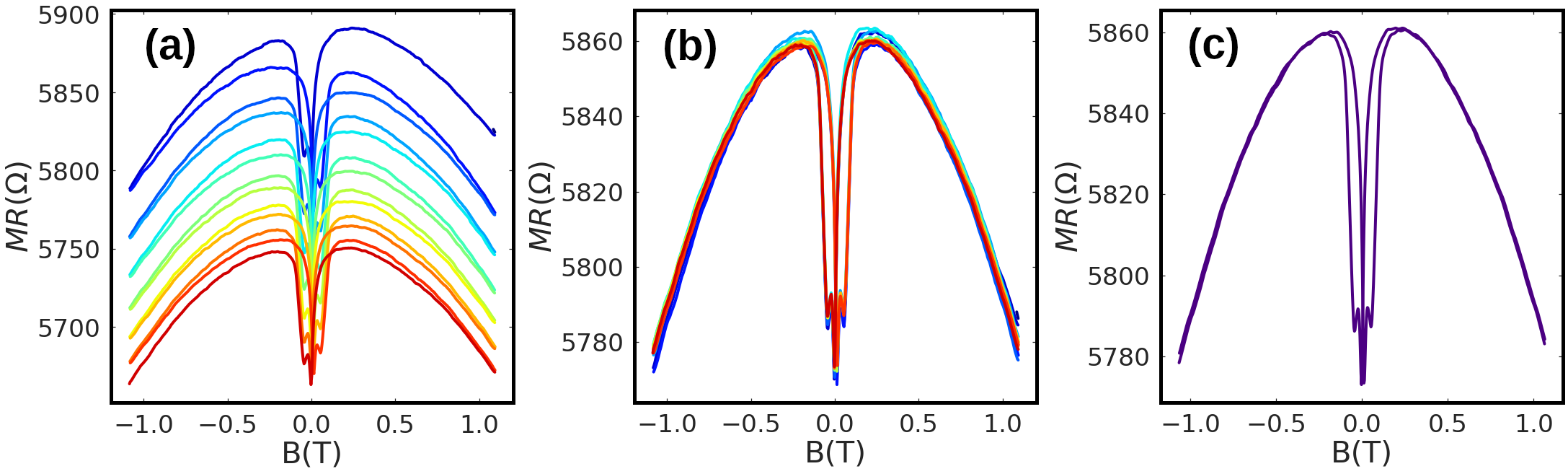}
	\caption{\label{DriftCorrectDemo} Demonstration of drift correction procedure for a longitudinal magnetoresistance trace taken at $\boldsymbol{\sim30}$mK and $\boldsymbol{V_g}$ = 200V. (a) shows raw data demonstrating a time dependent drift of magnetoresistance after a change in gate voltage. (b) drift corrected data showing good agreement after multiple traces. (c) shows the resulting trace after the repeated sweeps were interpolated and averaged together.}
\end{figure}

\begin{figure}[h]
\centering
\includegraphics[width=0.60\columnwidth]{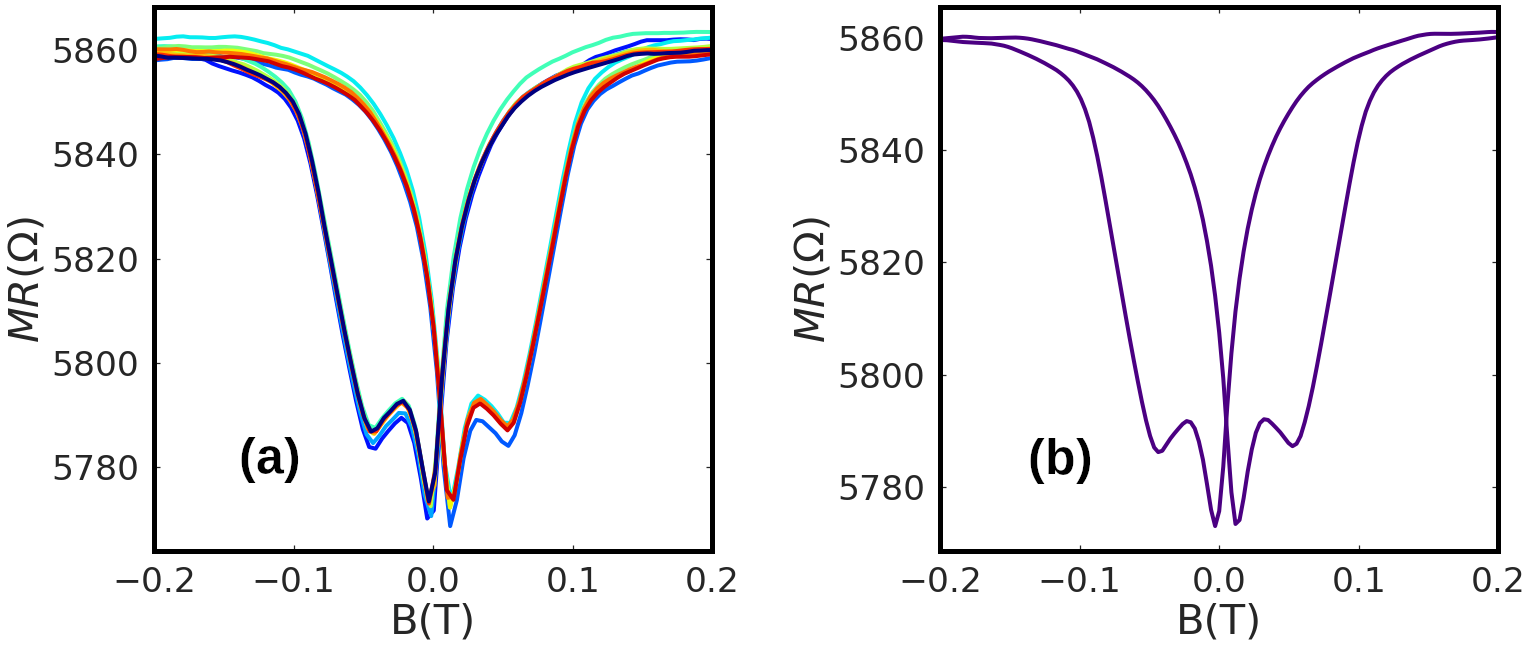}
\caption{\label{DriftCorrectLowB} Demonstration of low external magnetic field drift correction procedure. (a) shows drift corrected data showing good agreement after multiple traces, while (b) shows the resulting smoothed data, demonstrating the fidelity of the procedure.}
\end{figure}

\paragraph*{Drift Correction}
As mentioned in the main text, it is well known that complex oxide samples drift over time whenever changes are made to significant experimental parameters such as back gate voltage. This drift is persistent over the course of hours to days, so is accounted for in these reported data by subtracting off a time dependent logarithmic form as described by Biscaras et al [\textit{30}]. In Figure \ref{DriftCorrectDemo} (a) we show a trace presented as measured, with the resistance drift on top of the changes from magnetoresistance, taken over the course of several hours. Figure \ref{DriftCorrectDemo} (b) shows the corrected data after subtracting off the time dependent contribution described by Biscaras et al. of the form $R(t) = R_0 + A \log(t + C)$. Here we fit with the constants $R_0 = 1040$, $A = 250$, and $C = 10000$, where in this case $R_0$ is not the residual resistance but the shift necessary to recover the appropriate resistance at the start of the run, which is important for calculating percent magnetoresistance. The results shown in Figure \ref{DriftCorrectDemo} (b) are then interpolated along 751 evenly spaced values of magnetic field, and averaged together into one representative trace, as seen in Figure \ref{DriftCorrectDemo} (c). A zoomed in view of the center, showing the preservation of the hysteresis can be seen in Figure \ref{DriftCorrectLowB}. All longitudinal traces were subjected to this treatment, each with their own fitting parameters. Drift was also seen in the transverse magnetoresistance, but only for the lowest gate voltages, and these data were removed from the reported results because no reproducible Hall curve could be extracted. 
\newline{}
\newline{}

\begin{figure}[h]
\centering
\includegraphics[width=0.50\columnwidth]{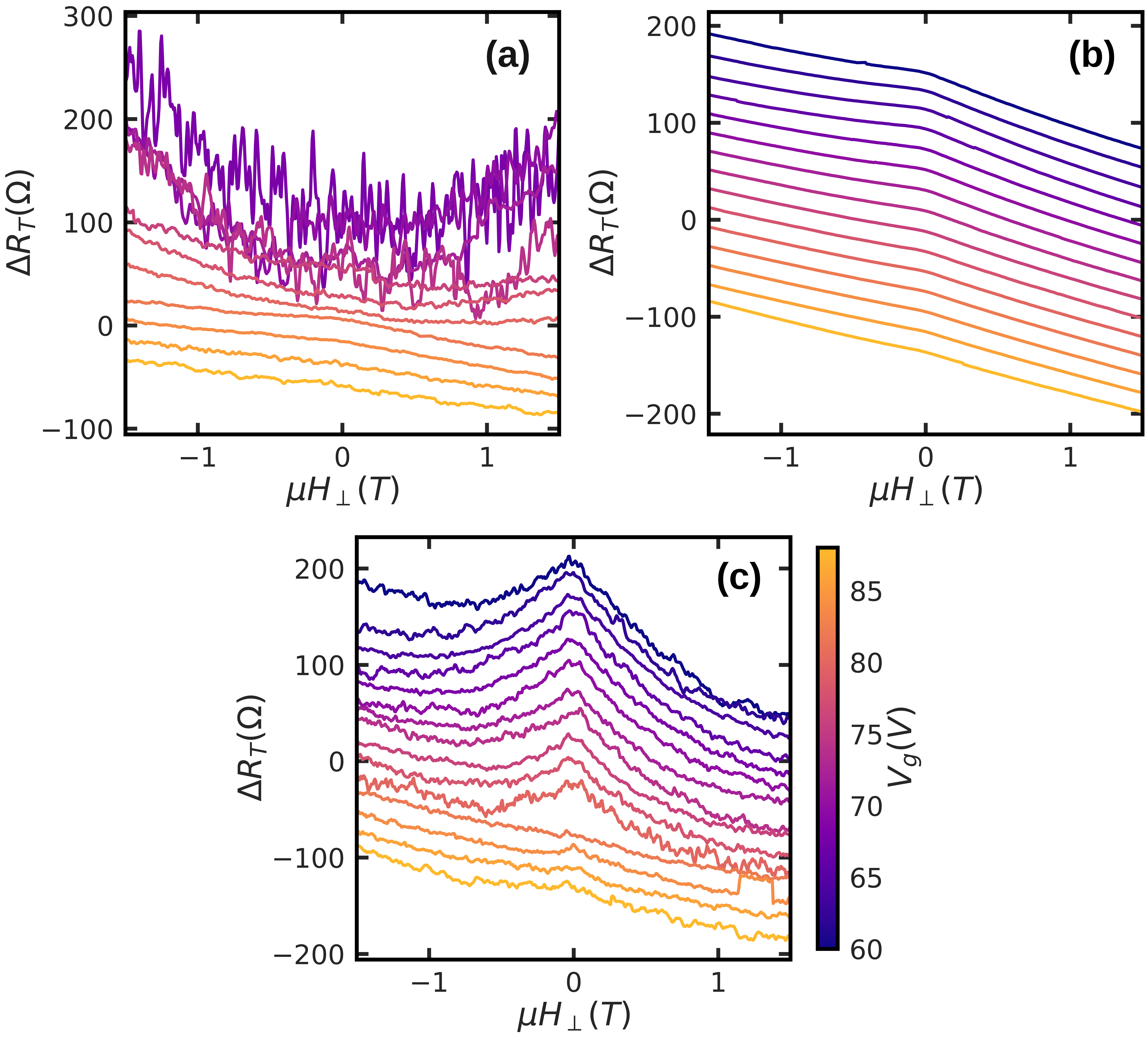}
\caption{\label{Detailed2K} Finer spaced traces of transverse magnetoresistance at 2K, taken around the inflection point at $V_{g} \approx 90V$ observed in Fig 1 of the Main Text. Data for the (a) (001), (b) (110), and (c) (111) show a marked change in behavior, most pronounced around $V_{g} \approx 80V$. These traces have been shifted by their zero field value and then with a uniform offset to help the presentation.}
\end{figure}

\paragraph*{Detailed Transverse Magnetoresistance at 2K}
To investigate the marked change in transport behavior around $V_{g} = 90 V$ a finer spaced measure of transverse voltage was taken. Data for the (001), (110), and (111) were taken simultaneously and presented in Fig. \ref{Detailed2K} (a-c), respectively. As was the case for data shown in Fig. 2 of the Main text, these data have the zero field value subtracted to facilitate comparison of the shape of the curves. The presented data were taken in sequence, without disturbing the measurement setup, in order of increasing $V_g$. A marked change of inflection is observed for the (001) and (111) samples below $V_{g} = 80 V$, with a gradual pronouncement of the central peak in the (110). Additionally, a significant increase in the background noise of the measurement is observed for these traces below the inflection point, largely due to an increase in the DC offset of Hall. Due to the drift inherent in the sample which was discussed in detail above, the exact behavior of the traces differ slightly from those same gate voltages reported in the Main Text. The effect is the same however, including the change from concave to convex in the (001) and (111). \newline{}

\begin{figure}
\centering
\includegraphics[width=0.8\columnwidth]{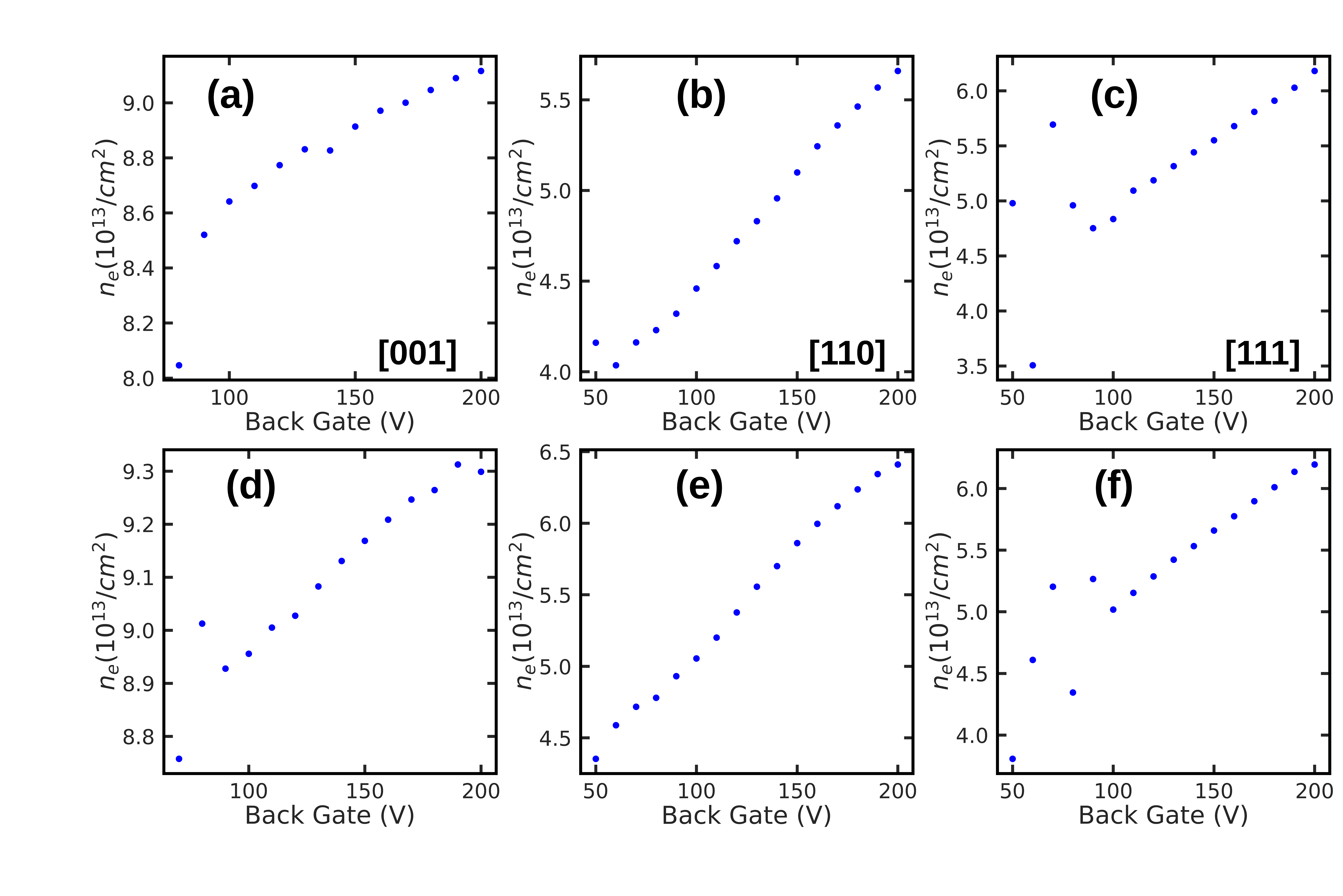}
\caption{\label{CarrierConc} Gate voltage dependent carrier concentration. (a-c) show the extracted carrier concentrations measured at $\sim$2.2K for the (001), (110), and (111) terminated samples respectively. (d-f) show the same carrier concentrations but measured at $\sim$5K.}
\end{figure}

\paragraph*{Carrier Concentration}
Carrier concentration and carrier mobility were calculated from the slope of the antisymmetric part of the Hall resistance $R_H = \Delta R / \Delta B$. Assuming a single band model gives the Hall coefficient as $R_H = -1 /n e$, where $e$ is the charge of the electron and $n$ is the areal density of the two dimensional conducting gas. The sign of the Hall coefficient, $R_H$, gives the sign of the carrier, which was confirmed to be electron-like. Our calculated carrier concentrations as a function of gate voltage are given in Figure \ref{CarrierConc} for each of the crystal terminations at measurement temperatures of $\sim$2.2K and $\sim$5K. Both sets of results rely on an assumed single band model and break down at low gate voltages due to competing effects that lower the precision of the measurement. Importantly, these calculated carrier concentrations agree well with other works that explored the nature of the superconductivity in this system [\textit{23}, \textit{31}], suggesting that the suppression of the superconductivity is not due to a deviation in the density of the 2DEG. From these data we also can extract carrier mobility, which ranges from a few tens of cm$^2$ V$^{-1}$ s$^{-1}$ to a few hundred, also in line with what other groups have reported. \newline{}

\begin{figure}[h]
\centering
\includegraphics[width=0.65\columnwidth]{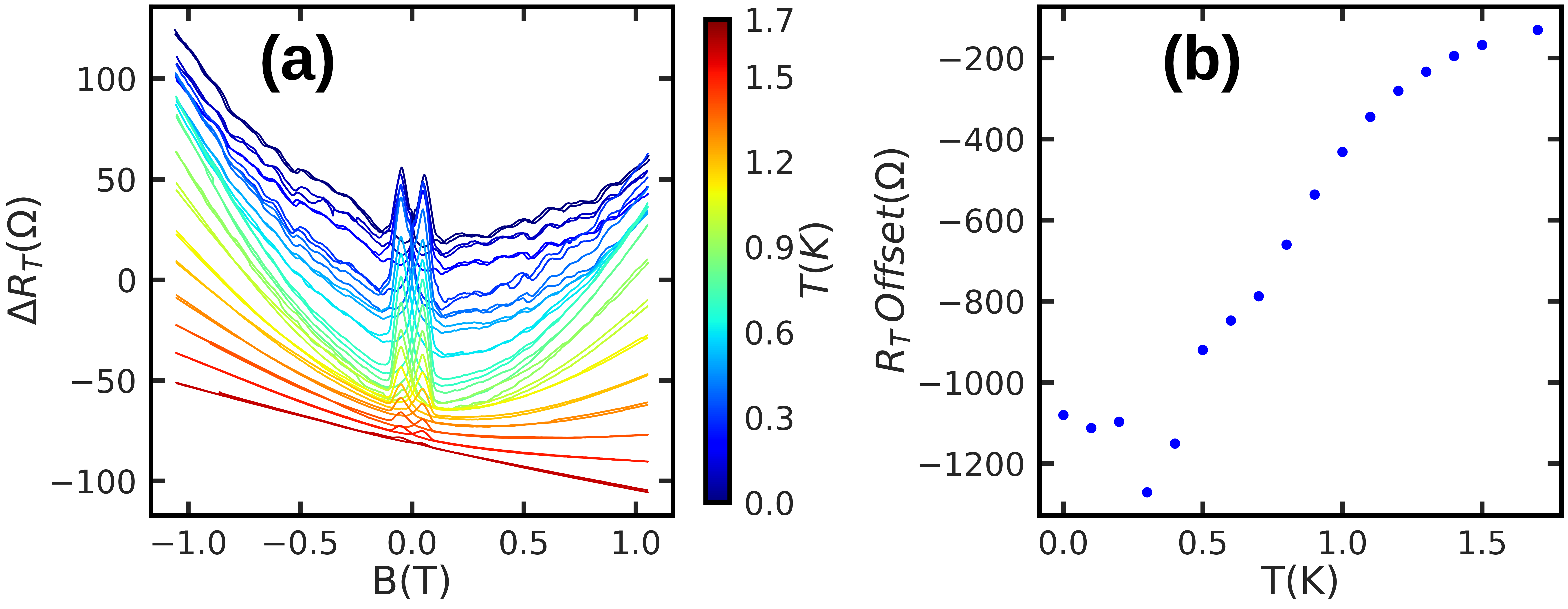}
\caption{\label{TempDepHysteresis} Temperature dependence of the low temperature Hall signatures. (a) shows the temperature dependence of the shifted Hall resistance for the (111) terminated sample from our base temperature of $\sim$30mK to the highest stable temperature of $\sim$1.7K. (b) shows the temperature dependent intrinsic anomalous Hall offset of the (111), described as the average measured resistance of the Hall curves along the field range shown in (a).}
\end{figure}

\paragraph*{Temperature Dependence of Hysteretic Magnetoresistance}
A detailed study of the samples' temperature dependent magnetoresistance was conducted to trace the evolution of the hysteresis with increasing temperature, and at a fixed gate voltage of $V_{g} = 200V$. These data revealed the temperature dependence of the three components of the magnetic transport signatures in the samples, as demonstrated in Figures \ref{TempDepHysteresis}. Figure \ref{TempDepHysteresis} (a) shows the shifted Hall traces for the (111) sample, which has two critical temperatures, one for the background oscillations, which are suppressed by $\sim$800mK, and one for the low field hysteresis, which is suppressed above $\sim$1.7K. These data have been shifted from their measured values as there is a large, field-independent intrinsic Hall contribution which is subtracted off. They are also spread out in even intervals of resistance for clarity. The subtracted intrinsic Hall shift is plotted as a function of temperature in Figure \ref{TempDepHysteresis} (b). The measured data were limited by the available temperature range of the probe, but show a saturation as we approach the other reported results at $\sim$2.2K. \newpage

\newpage

\subsection*{Computational Methods}
\paragraph*{Surface 2DEG and Spin Texture}
KTaO$_{3}$ belongs to spacegroup 221 with lattice parameter $a=4.03$ \angstrom. The lattice parameters and atomic positions are taken from the Materials Project [\textit{41}]. All first principles calculations are based on density functional theory and were performed using the Quantum ESPRESSO software package [\textit{42}]. The calculations utilize the generalized gradient approximations (GGA) of Perdew-Burke-Ernzerhoff (PBE) [\textit{43}]. Spin-orbit coupling is included in the calculations.

\begin{figure*}[hb]
    \centering
    \includegraphics[width=0.75\columnwidth]{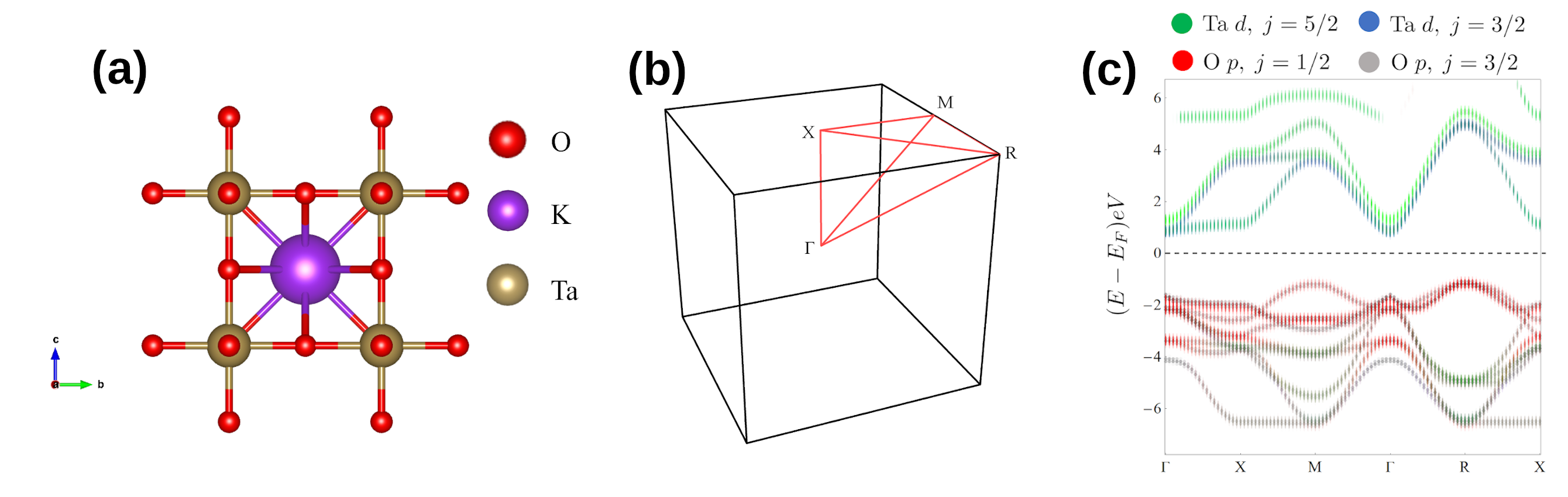}
    \caption{\label{BandStructure}(a) Primitive unit cell of bulk KTaO$_{3}$ belonging to spacegroup 221. (b) Brillouin zone of primitive unit cell with high-symmetry locations and path labeled. (c) Momentum space resolved density of states along the high-symmetry path labeled in (b), demonstrating that the conduction bands are primarily composed of the Ta $d$-orbitals.}
\end{figure*}

For bulk calculations, a plane-wave cutoff of 60 Ry is used and the primitive unit cell is sampled with a Monkhorst k-mesh of $9 \times 9 \times 9$. Spin-orbit coupling is included in the calculations. The calculated band structure resulting from a bulk unit cell is shown in Fig. \ref{BandStructure}.

\begin{figure}[H]
    \centering
    \includegraphics[width=0.7\columnwidth]{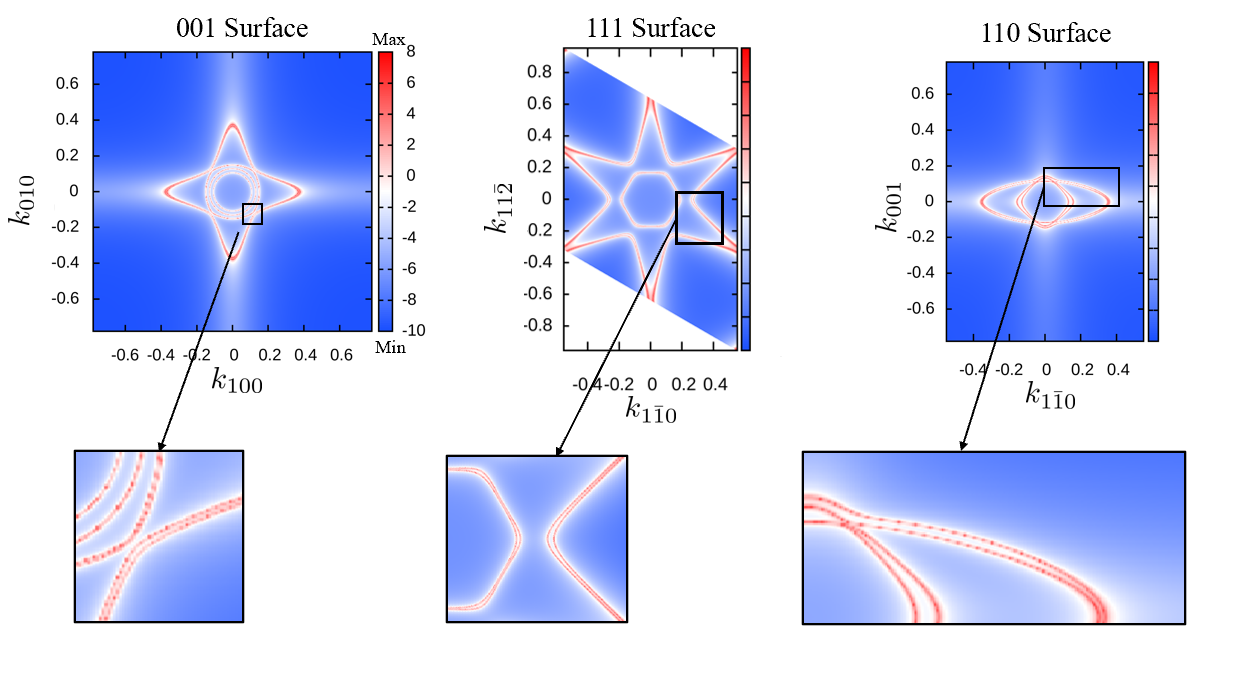}
    \caption{Momentum space resolved spectral weight for three distinct surface terminations of KTaO$_{3}$ illustrating Fermi surface of 2DEG. Inset shows zoomed picture of Fermi surface to demonstrate the $k$-dependent Rashba splitting.}
    \label{fig:SurfaceStates}
\end{figure}

\begin{figure*}
    \centering
    \includegraphics[width=0.85\columnwidth]{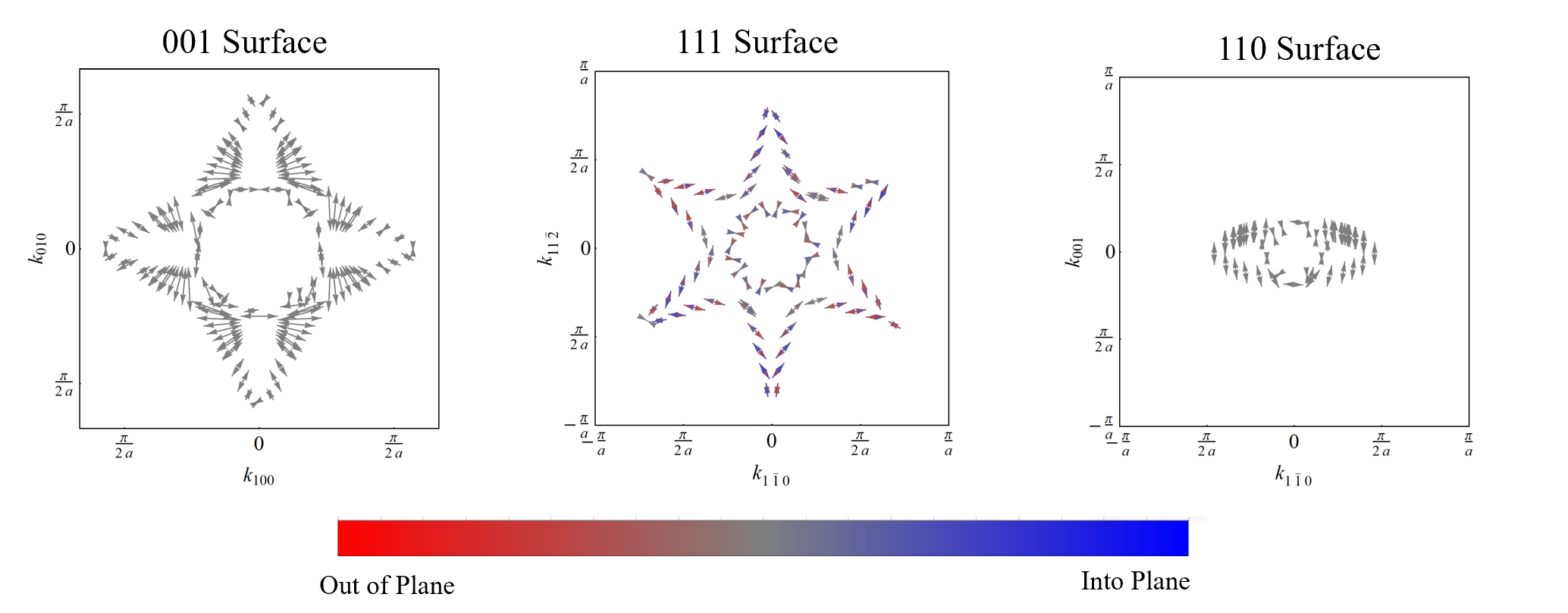}
    \caption{Spin texture for Fermi surface of 2DEG on three distinct terminations of KTaO$_{3}$. Color is used to illustrate spin texture perpendicular to the plane. Only the (111) surface displays a non-vanishing spin-texture perpendicular to the plane. Yellow outline indicates the Fermi surface.}
    \label{fig:SpinTexture}
\end{figure*}

In order to calculate the surface spectrum and spin-orbit texture, a Wannier tight-binding model is generated from the $p_{x,y,z}$ orbitals of the O atoms and the $d_{xy,yz,zx}$ orbitals of Ta using the Wannier90 software package [\textit{44}]. Following Ref. [\textit{25}], the 2DEG at the surface can be modeled through the introduction of a potential well at the surface in order to avoid the explicit introduction of symmetry breaking terms. The magnitude of this potential well is the only parameter tuned in the calculations and is fitted such that the results are in alignment with those of Ref. [\textit{25}]. The surface spectra and spin-orbit texture is then calculated using the WannierTools software package [\textit{45}]. The results of the surface spectra calculations for the three crystal terminations are shown in Fig. \ref{fig:SurfaceStates}. Our calculations indicated a maximum Rashba coefficient, $\alpha_{R} \approx$ 2 meV\angstrom, for the (111) and (110) surfaces, in line with what is found by Bruno et. al [\textit{25}] and $\alpha_{R} \approx$ 1 meV\angstrom \,for the (001) surface. Similarly, calculations of the spin texture reveal an out of plane contribution for the (111) surface which is absent in the (001) and (110) samples, as shown in Fig. \ref{fig:SpinTexture}. These results also agree with those published by Bruno et. al [\textit{25}]. 
% \newline{}

\paragraph*{Presence of Finite Surface Berry-Curvature Density}
The results shown in Fig. 6 of the main text were calculated for the (111) surface utilizing the same Wannier tight binding model with a surface potential well generated to study the 2DEG spectra. The Berry curvature is then calculated discretizing the Brillouin zone into a 200×200 grid of plaquettes. The wavefunction of all occupied states is then parallel transported around the plaquette following the procedure put forth by Fukui et. al [\textit{46}]. 

\begin{figure}[h]
    \centering
    \includegraphics[width=0.8\columnwidth]{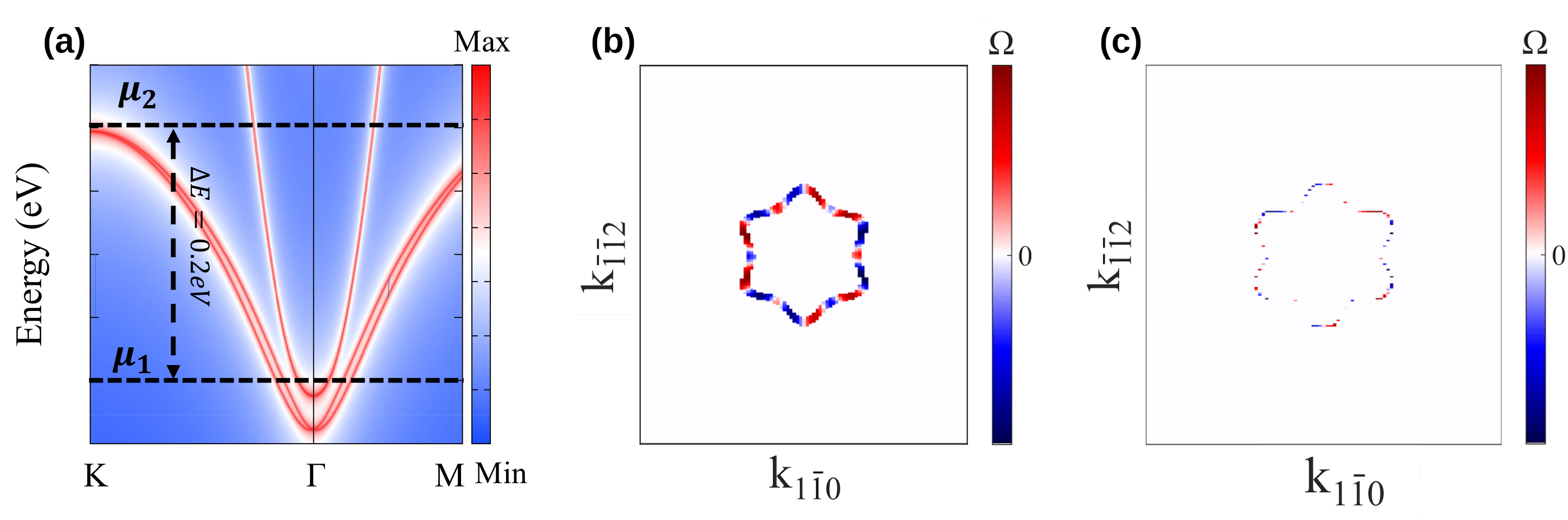}
    \caption{(a) Spectral density on the (111) surface along high-symmetry path in the Brillouin zome. Berry curvature density is measured at two filling fractions, (a) $\mu_{1}$ and (b) $\mu_{2}$.  The cut at ($\mu_{2}$) demonstrates the highly diminished Berry curvature density at larger filling fractions.}
    \label{fig:SecondCut}
\end{figure}

As a supplement to Fig. 6 of the Main Text we have included the second cut 200meV above the filling fraction shown Fig. 6 (b), shown in Fig. \ref{fig:SecondCut}. This re-arraingement shows the effect of increasing filling: a decrease in the Rashba splitting leads to a vanishing Berry curvature. This is most obvious in the lack of contribution in Fig. \ref{fig:SecondCut} (b) for the bands closest to the $\Gamma$-point, and for the suppression of the Berry curvature in Fig. \ref{fig:SecondCut} (c).
\newline{}

\begin{figure}[h]
    \centering
    \includegraphics[width=0.65\columnwidth]{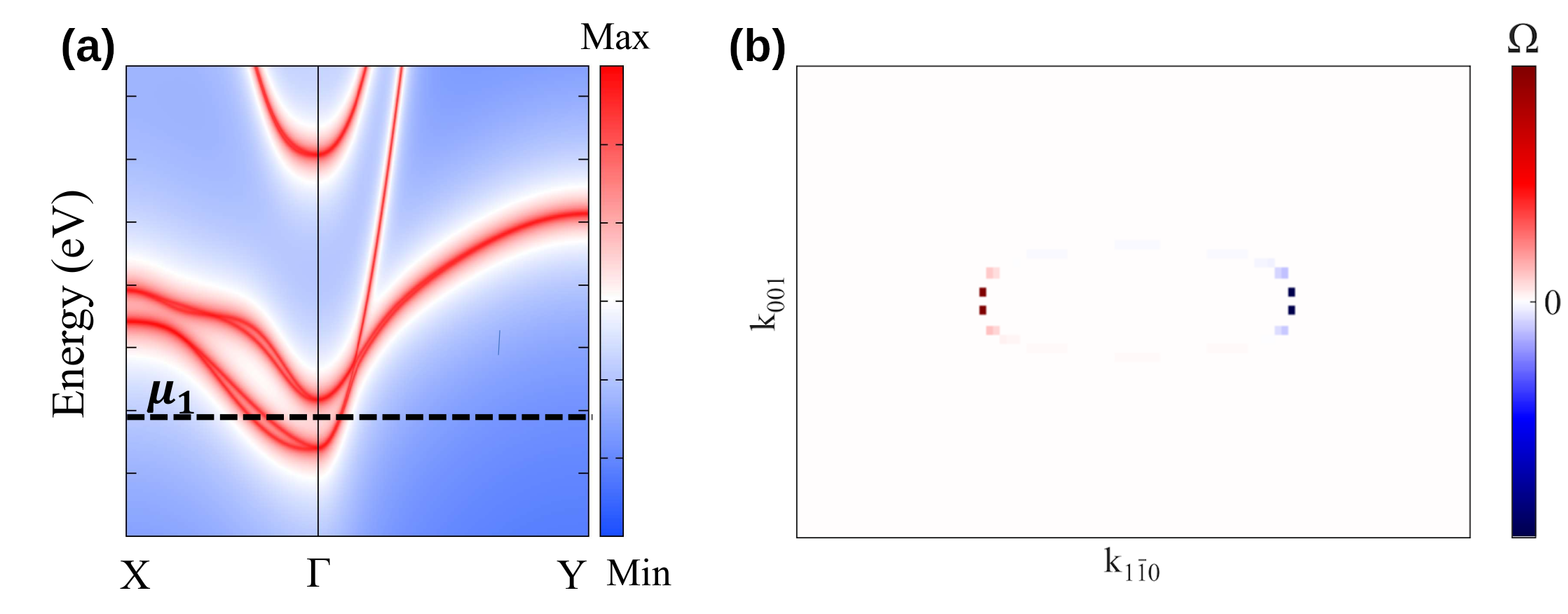}
    \caption{((a) Spectral density on (110) surface along high-symmetry path in surface Brillouin zone. The lowest lying bands exhibit large effects from Rashba splitting at low energy upon application of surface potential well. Berry curvature density is measured at filling fraction $\mu_{1}$. (b) Berry curvature density at low filling fraction ($\mu_{1}$). The finite Berry curvature arises when only the lowest lying Rashba split band is occupied. A larger Rashba splitting along the $\Gamma$ - X line leads to a larger Berry curvature density in this region.}
    \label{fig:110_Curvature}
\end{figure}

We also calculated Berry curvature for the (001) and (110) surfaces, once again along high-symmetry paths in the Brillouin zone. The (001) did not show any appreciable Berry curvature density, due to its small Rashba splitting shown in Fig. \ref{fig:SurfaceStates}. A more detailed spectral density for the (110) is shown in \ref{fig:110_Curvature} (a) with a filling fraction cut ($\mu_{1}$). The small, but finite, Berry curvature density measured along ($\mu_{1}$) is shown in Fig. \ref{fig:110_Curvature} (b). In comparison to even the high energy cut in Fig. \ref{fig:SecondCut} (c), the resulting Berry curvature density in the (110) is small, in agreement with our experimental observations.
\newline{}
 
\paragraph*{Emergent Magnetism: The (001) Surface}
To examine the presence of emergent surface magnetism, leading to a finite zero field charge Hall effect, a supercell is constructed of size $2 \times 1 \times 6$. The (001) surface is perpendicular to the $c$-axis of the supercell and a minimum of 20 \angstrom \, of vacuum is added along the $c$-direction. We first focus on the (001) surface as it displays a zero field charge Hall conductivity while limiting computational expense. The terminations of the supercell are made symmetric. Multiple relaxations are performed, one for each oxygen vacancy that can be formed at the top layer. The relaxation calculations are performed using a plane-wave cutoff of 60 Ry, a charge cutoff of 550 Ry and the cell is sampled with a Monkhorst k-mesh of $2 \times 4 \times 1$, magnetization is considered in relaxation calculations and the bottom three layers are held fixed. Following relaxation the lowest energy structure is selected for a self consistent calculation. The self-consistent calculation is performed utilizing a Monkhorst k-mesh of $4 \times 8 \times 1$ followed by a non-self consistent calculation with a Monkhorst k-mesh of $8 \times 16 \times 1$. Following this, the density of states was calculated to determine spin-polarization and charge distribution. The simulation estimates a total magnetic moment of $~0.56$ Bohr magneton for the full supercell with two surface Ta atoms. This result is consistent with the relative magnetude of the experimental results for the (001) sample. The process was repeated with no oxygen vacancies to ensure the system converged to a non-magnetic state. 

\begin{figure*}[h]
\centering
\includegraphics[width=0.55\columnwidth]{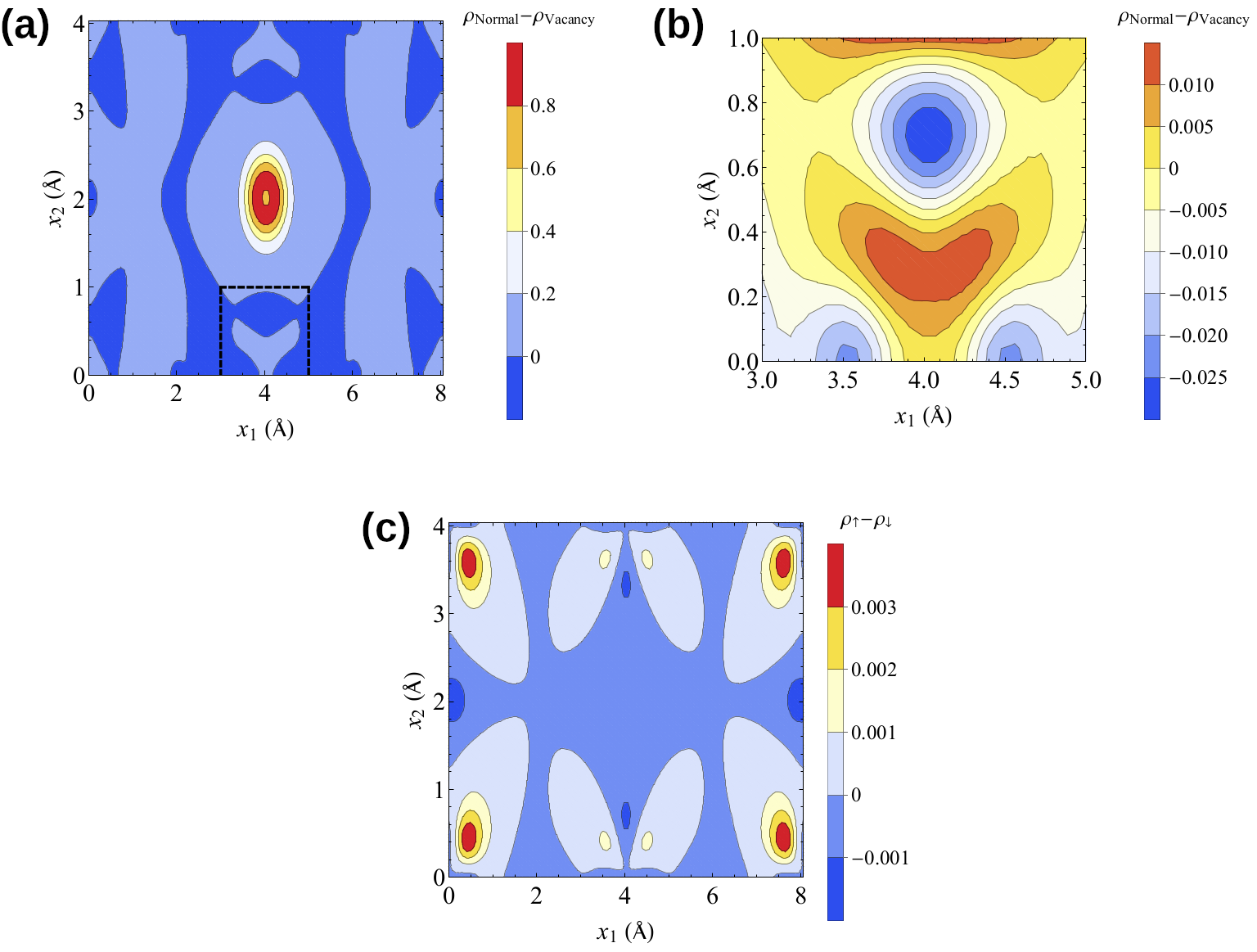}
\caption{\label{001SurfaceDFT}(a) Difference in charge density on the surface for the supercell with an oxygen vacancy at the center ($\rho_{vacancy}$) and without ($\rho_{normal}$). The dashed box outlines the location of the Ta atom nearest to the vacancy. The location of the oxygen vacancy at the center is evident by the increased values in that region. (b) The charge density inside the dashed box shown in (a). (c) Spin polarization on the surface of the supercell with an oxygen vacancy. Any significant contribution to spin-polarization is localized to the Ta atoms.}
\end{figure*}

The partial charges of each atom were then compared for the systems with and without the charge vacancy. The atom which displays most significant change in partial charge is the Ta atom closest to the oxygen vacancy on the surface with a change in charge of approximately $0.5 e$. This is visible in the density plot shown in Fig. \ref{001SurfaceDFT}(a).   \newline{}

\begin{figure}[H]
\centering
\includegraphics[width=0.5\columnwidth]{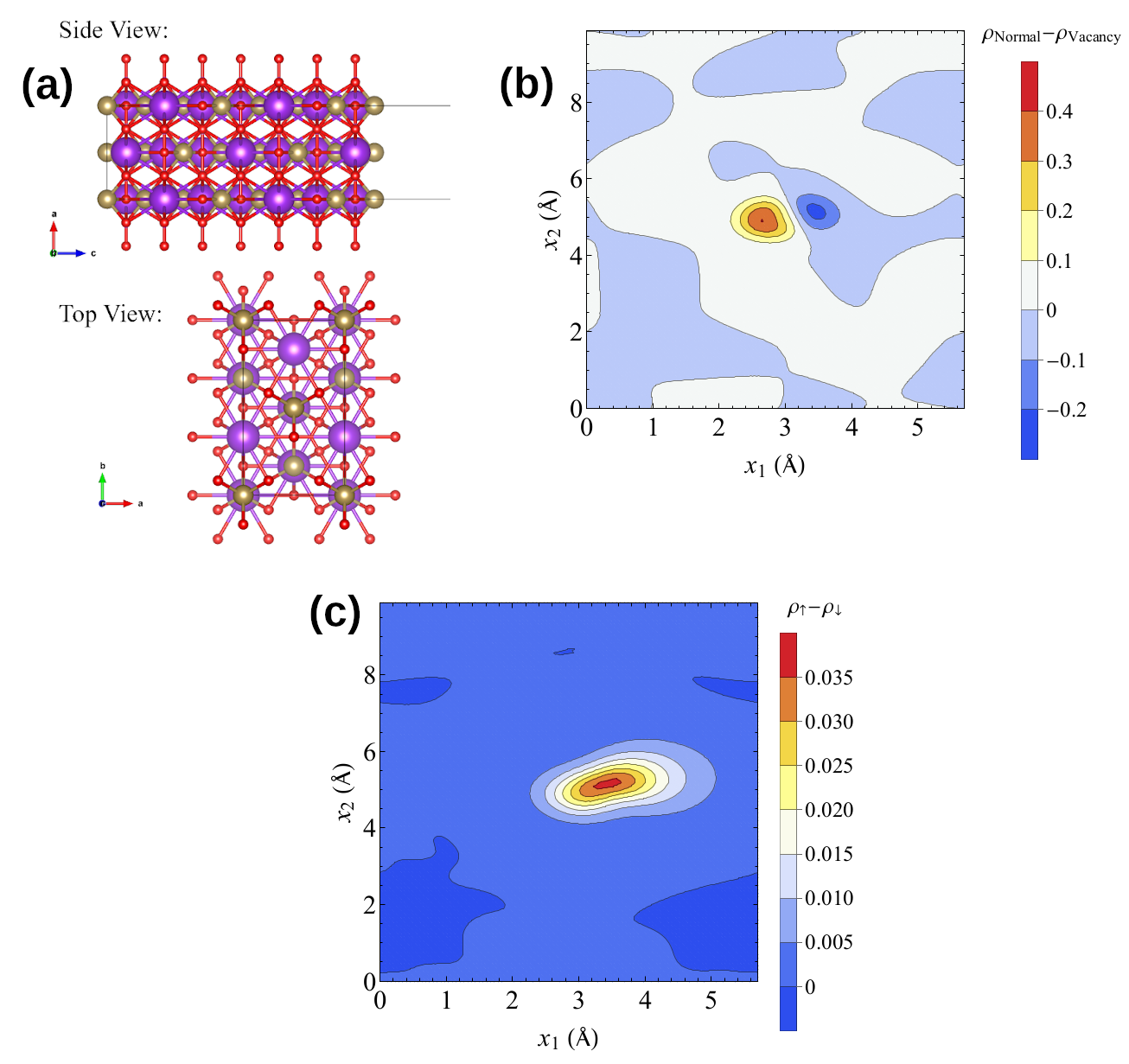}
\caption{\label{111SurfaceDFT}(a) Supercell used in simulating the (111) surface. (b) Difference in charge density on the surface for the supercell with an oxygen vacancy at the center ($\rho_{vacancy}$) and without ($\rho_{normal}$). Unlike the (001) surface the oxygen vacancy is offset from the plane containing the Ta atoms, removing the large differential seen in Fig. \ref{001SurfaceDFT}. (c) Spin polarization on the surface of the supercell with an oxygen vacancy. Any significant contribution to spin-polarization is localized to the Ta atoms.}
\end{figure}

\paragraph*{Emergent Magnetism: The (111) Surface}
To study the emergent surface magnetism on the (111) surface, we construct a supercell with the $c$-axis along the [111] direction. It is further constructed such that the surface terminations are equivalent and there exists two Ta atoms on the surface. Again 20 \angstrom \, of vacuum is added along the $c$-direction to limit finite size effects. The supercell is visible in Fig. \ref{111SurfaceDFT} (a). The relaxation procedure listed above for a single oxygen vacancy on the surface is repeated using identical energy cutoffs and k-point grids. The simulation estimates a total magnetic moment of $~6.67$ Bohr magneton for the full supercell with two surface Ta atoms. Importantly, the total magnetic moment has increased significantly from the (001) surface. We should note that the supercell used to model the (111) surface has 71 atoms as opposed to 35 atoms for the (001) surface, however, this still represents a significant qualitative increase and is in accordance with an increased zero-field charge Hall conductivity on the (111) surface. The process was repeated with no oxygen vacancies to ensure the system converged to a non-magnetic state.

The partial charges of each atom were then compared for the systems with and without the charge vacancy. The atom which displays most significant change in partial charge is the Ta atom on the surface, closest to the oxygen vacancy with a change in charge of approximately $0.5 e$. This is visible in the density plot shown in Fig. \ref{111SurfaceDFT} (b).